\documentclass{jfm}
\usepackage{graphicx}
\usepackage{epstopdf, epsfig}
\usepackage{caption}
\usepackage{subcaption}
\usepackage{amsmath}
\usepackage{amssymb}
\usepackage{wasysym}
\usepackage{upgreek}
\usepackage{multirow}
\usepackage[table,xcdraw]{xcolor}
\usepackage{booktabs}
\usepackage{gensymb}
\usepackage{rotating}
\usepackage{soul}

\shorttitle{Scaling analysis of the swirling wake of a porous
disc}
\shortauthor{E. Fuentes Noriega and N. Mazellier}

\title{Scaling analysis of the swirling wake of a porous
disc: application to wind turbines}

\author{E. Fuentes Noriega\aff{1}, \and N. Mazellier\aff{1}\corresp{\email{nicolas.mazellier@univ-orleans.fr}}}

\affiliation{\aff{1}University of Orléans, INSA-CVL, PRISME, EA 4229, 45072 Orléans, France}

\begin{document}

\maketitle

\begin{abstract}

\noindent We report a comprehensive study of the wake of a porous disc, the design of which has been modified to incorporate a swirling motion at an inexpensive cost. The swirl intensity is passively controlled by varying the internal disc geometry, i.e. the pitch angle of the blades. A swirl number is introduced to characterise the competition between the linear (drag) and the azimuthal (swirl) momentums on the wake recovery. Assuming that swirl dominates the near wake and non-equilibrium turbulence theory applies, new scaling laws of the mean wake properties are derived. To assess these theoretical predictions, an in-depth analysis of the aerodynamics of these original porous discs has been conducted experimentally.  It is found that at the early stage of wake recovery, the swirling motion induces a low-pressure core, which controls the mean velocity deficit properties. The measurements collected in the swirling wake of the porous discs support the new scaling laws proposed in this work. Finally, it is shown that, as far as swirl is injected in the wake, the characteristics of the mean velocity deficit profiles match very well those of both lab-scale and real-scale wind-turbine data extracted from the literature. Overall, our results emphasise that by setting the initial conditions of the wake recovery, swirl is a key ingredient to be taken into account in order to faithfully replicate the mean wake of wind turbines.

\end{abstract}


\section{Introduction}\label{sec:introduction}

\noindent To tackle the challenges posed by climate change, global policies increasingly advocate for a significant rise of renewable energy within the energy mix. In this context, the wind energy sector has shown an exceptional growth, raising problems related to the sizing, positioning and operation of wind turbines. The trade-off between technical constraints and resource availability usually results in wind turbines being densely grouped in clusters known as wind farms, whose efficiency largely depends on the so-called wake interactions. Wind power extracted by a wind turbine is proportional to its drag coefficient $C_D$ defined as follows

\begin{equation}
	C_D = \frac{F_D}{\frac{1}{2}\rho U_\infty^2 A},
	\label{eq:drag_coeff_def}
\end{equation}

\noindent where $F_D$ is the drag force experienced by the body, $\rho$ is the air density, $U_\infty$ is the incoming velocity and $A$ is the frontal area of the body. Since $C_D$ is intimately linked to wake recovery, accurately predicting the development of wind turbine wakes is crucial for optimising wind farm operation. However, the number of degrees of freedom characterising this problem is so vast that it is currently impossible to solve it with high-fidelity numerical tools. To overcome this issue, simplified approaches are required. A popular example of this strategy is the actuator disc concept first introduced by \cite{rankine1865mechanical}, which assimilates the wind turbine to a porous medium surrogate across which a pressure drop can be tuned to match the drag coefficient \citep{van2015rotor}.

\noindent At the laboratory scale, the actuator disc concept has been implemented using porous discs to mimic isolated wind turbines \citep[see e.g.,][]{Aubrun2013, Howland2016} or wind farms \citep[see e.g.,][]{bossuyt2017wind, Camp2016, stevens2018comparison}. As a starting point, the seminal work of \cite{castro1971perforatedplates} on perforated flat plates, later on extended by \cite{steiros2018drag} emphasise the attractiveness of this analogy reducing the problem complexity to a single physical parameter: the porosity $\beta$, which represents the ratio of empty volume to total volume. Based on potential flow theory and conservation laws, \cite{steiros2018drag} derived a relationship between $C_D$ and $\beta$, which was successfully validated against experimental measurements for any porosity value. This means that tuning porosity is a simple and efficient way to match the drag coefficient of a wind turbine. For instance, \cite{sforza1981three} studied experimentally the wake generated by porous discs within various operating conditions. \cite{Aubrun2013} conducted an experimental survey comparing the wake generated by a lab-scale rotating wind turbine to that produced downstream a porous disc with adjusted porosity. Analysing mean velocity and turbulence intensity profiles, these authors concluded that beyond 3 rotor diameters both wakes behave similarly. Identical conclusions were reached by \cite{stevens2018comparison} who investigated the influence of the wind turbine model incorporated in Large Eddy Simulations for both a single wind turbine and a wind farm. Comparing their results to the experimental database reported by \cite{chamorro2011turbulent}, these authors showed that the actuator disc model is well adapted to capture the main features of the mean wake.

\begin{figure}
    \centering
    \includegraphics[width=0.96\textwidth]{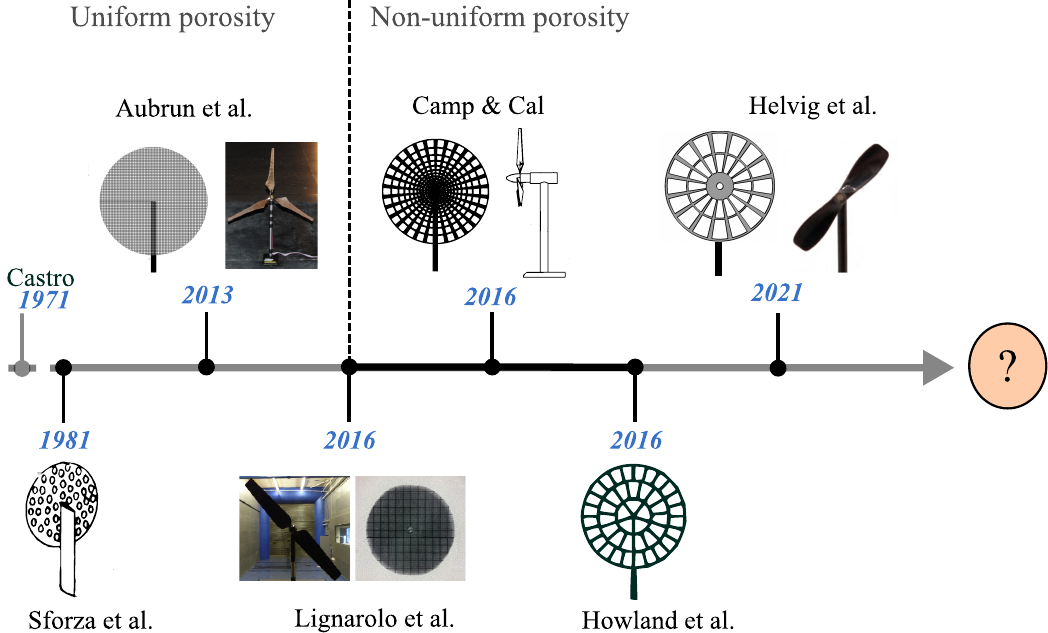}
    \caption{Schematic diagram showing the chronology of the porous disc model used as a wind turbine surrogate for wind tunnel experiments and corresponding references.}
    \label{fig:PD_history}
\end{figure}

As depicted in Fig. \ref{fig:PD_history}, the porous disc model has shown an evolution throughout the years marked by a shift in focus towards porosity distribution - from models with uniform porosity \citep{sforza1981three,Aubrun2013,Lignarolo2016exp} to those featuring non-uniform porosity distribution \citep{Camp2016,Howland2016,helvig_2021}. The progression toward increasingly intricate designs, with the aim of more accurately mirroring the blades of a wind turbine, is evident. Recently, \cite{Aubrun2019} and \cite{vinnes2023actuator} performed a detailed comparison of both types of porosity distribution in different facilities. The authors found discrepancies from flow to flow that they attributed to variations in the initial conditions of the wake, which are known to have a very long-range influence \citep{bevilaqua1978turbulence,wygnanski1986large}. This observation resonates with the results reported by \cite{stevens2018comparison} who showed that by adding more details to the initial conditions of the actuator disc model (like the nacelle, hub and mast) greatly increased its fidelity. This tends to show that by restricting the design of the actuator disc concept to a single parameter, porosity, the relevance of this model is likely to remain limited.

\cite{Camp2016} pointed out that although the porous disc model can closely approximate most statistics of the mean flow, it inherently lacks the ability to replicate swirl, which is a defining feature in the near wake of wind turbines \citep{PorteAgel2019}. This additional motion comes from the rotation of the blades, which confer an angular momentum $G_0$ to the wake. Knowing that swirl plays a crucial role in areas like combustion \citep{masri2004compositional} and geophysics \citep{moisy2011decay}, its omission in the design of actuator disc may result in oversimplified wind turbine surrogate. In fact, Professor \cite{joukowsky1912vortex} had already emphasised the relevance of rotation in screw vortex systems like propellers, helicopters and wind turbines (readers interested in Joukowsky's legacy to the development of rotor theory are referred to \cite{okulov2015rotor} and \cite{van2015rotor}. Introducing the swirl number $F_D \mathcal{L}/G_0$, where $\mathcal{L}$ is a characteristic length scale (typically the rotor diameter), \cite{reynolds1962similarity} investigated the self-similar solutions by considering two asymptotic cases: linear momentum (drag) dominated flows, i.e., $F_D \mathcal{L}/G_0 \gg 1$, and angular momentum (swirl) dominated flows, i.e., $F_D \mathcal{L}/G_0 \ll 1$. While the latter regime is reminiscent of wakes behind self-propelled bodies \citep{chernykh2005swirling}, the former regime corresponds to the framework in which most popular wake recovery models were established \citep{jensen1983note,frandsen2006analytical,Bastankhah2014}. Recently, \cite{holmes2022impact} estimated a range of swirl number values for real wind turbines using the results from NREL's FAST8 model \citep{bortolotti2019iea}. Swirl numbers reaching values up to 0.25, i.e., $\mathcal{O}(1)$, were found, meaning that the influences of both linear and angular momentum are significant and cannot be disregarded. In this scenario, the consideration of swirl becomes indispensable in accurately predicting wind turbine wake development, marking the core focus of this work.

\noindent The objective of this investigation is twofold and is based on the following two questions: Can we inject swirl in a simple, straightforward manner with an adjustable parameter to the porous disc analogy? Can we predict the development of a swirling wake and characterise the swirling wake to verify it? The outline of the paper is structured to answer these questions as follows: \S 2 examines the theoretical implications of considering swirl in the development of a turbulent axisymmetric wake with particular attention to the similarity analysis; \S 3 describes the porous disc design process to passively include swirl and presents the experimental set-up. The aerodynamic performances of the proposed porous disc is assessed in \S 4, which is completed by a mean wake survey in \S 5 where a scaling analysis is conducted. Conclusions are drawn in \S 6 along with some perspectives.

\section{Self-similarity analysis of the mean swirling wake}
\label{sec:similarity}

\begin{figure}
                \centering
                \includegraphics[width=0.8\textwidth]{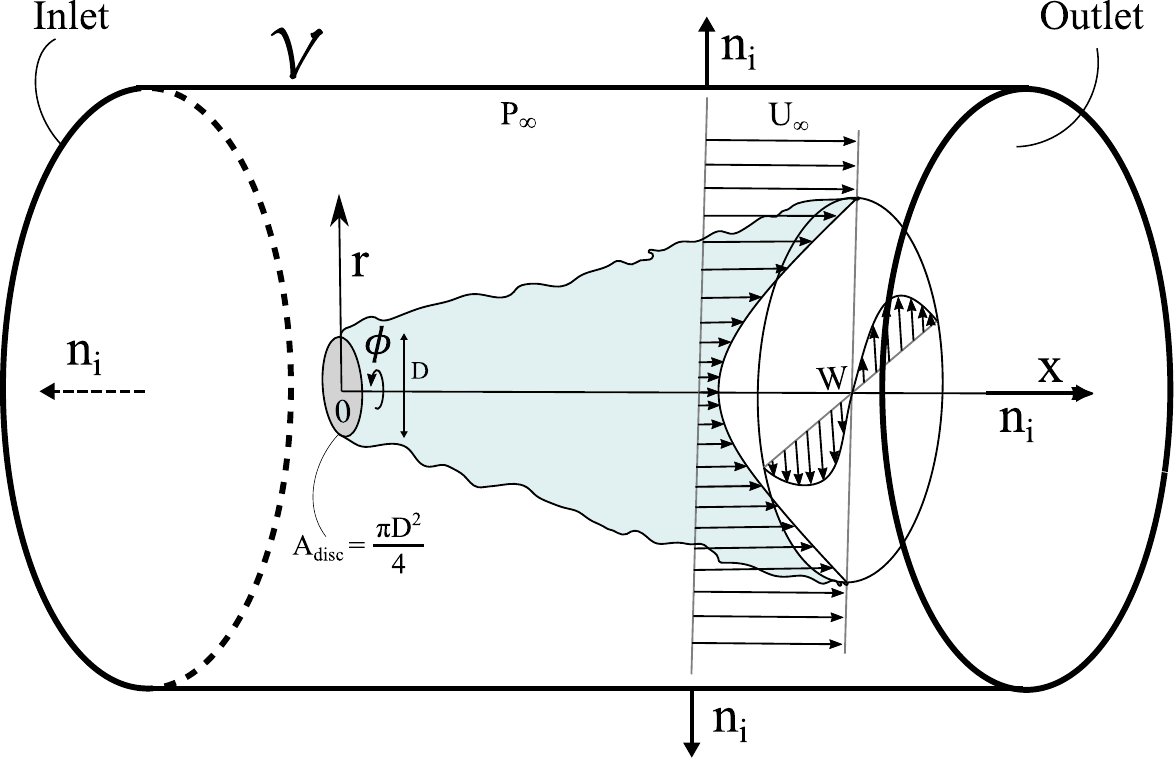}
                \caption{Schematic diagram of the axisymmetric swirling wake generated by an actuator disc contained in a control volume $\mathcal{V}$. $U$ and $W$ are the mean streamwise and azimuthal components, respectively. $P_\infty$ is the free-stream static pressure and $n_i$ are the outbound vectors normal to the surfaces. }
                \label{fig:momentum_budget_CV}
            \end{figure}

\noindent In this section, the scaling laws of the main parameters featuring a swirling wake are derived based on a self-preserving approach. To this end, simplified conservation laws are first established. Then, different scenarios are explored depending on the state of the dissipation rate of turbulent kinetic energy. As illustrated in Fig. \ref{fig:momentum_budget_CV}, we consider an axisymmetric wake generated by an actuator of diameter $D$ centred at the origin of the cylindrical coordinates system ($x$, $r$, $\phi$). The swirling motion of the wake is triggered by an angular momentum $G_0$ injected to the flow at the actuator disc location.  The wake is characterised by its velocity components $u$, $v$ and $w$ along the streamwise ($x$), the radial ($r$) and the azimuthal ($\phi$) directions, respectively. The actuator disc is subjected to an inflow characterised by a free stream velocity $U_\infty$. For the remainder of the paper, the symbol $\bullet^\star$ denotes normalised quantities using $U_\infty$ and $D$ as characteristic scales for velocities and lengths, respectively.

\subsection{Governing equations and relevant simplifications}

\noindent Since the theoretical framework describing the development of axisymmetric free shear flows has been established by previous works \citep[see e.g.][]{Townsend1976,George89,Johansson2003,Shiri2008}, we will only give a brief review of the main equations on which the self-similar analysis is based. Under the boundary-layer approximation (i.e $\partial/\partial r \gg \partial/\partial x$), the mean momentum transport equation along the streamwise direction is the following

\begin{equation}
U\frac{\partial U}{\partial x} + V\frac{\partial U}{\partial r} = - \frac{1}{r}\frac{\partial (r\overline{u'v'})}{\partial r} + \frac{\partial}{\partial x} \left\{  \overline{v'^2}-\overline{u'^2} + \int_r^\infty\frac{W^2 + (\overline{w'^2}-\overline{v'^2})}{r'}dr' \right\},
\label{eq:xmomentum_reduced_appa}
\end{equation}

\noindent where $U$, $V$ and $W$ are the mean velocity components, while $u'$, $v'$ and $w'$ represent the turbulent fluctuations along the streamwise ($x$), the radial ($r$) and the azimuthal ($\phi$) directions, respectively. The symbol $\overline{\bullet}$ stands for the ensemble averaging operator. Note that in this expression, the streamwise pressure gradient is inferred via the mean momentum transport equation along the radial direction leading to the term in brackets \citep{Shiri2008}. Equivalently, the mean momentum transport equation along the azimuthal direction reads        
            
 \begin{equation}           
U\frac{\partial W}{\partial x} + V\frac{\partial W}{\partial r} +\frac{V W}{r}   =   -\frac{\partial \overline{v'w'}}{\partial r} -2\frac{\overline{v'w'}}{r} - \frac{\partial \overline{u'w'}}{\partial x}. \label{eq:phi_momentum}            
\end{equation}

\noindent Integrating Eq. (\ref{eq:xmomentum_reduced_appa}) over a control volume $\mathcal{V}$ encompassing the actuator disc (see Fig. \ref{fig:momentum_budget_CV}) leads to the following expression

\begin{multline}
C_D = \frac{2 F_D}{\rho U_\infty^2 A_{\mbox{disc}}} =  16 \underbrace{\int_0^\infty U^*\Delta U^* r^*dr^*}_{\text{I}} - 16 \underbrace{\int_0^\infty\left[\overline{u'^2}^* - \frac{\overline{w'^2}^* + \overline{v'^2}^*}{2}\right]r^*dr^*}_{\text{II}} + \\
16 \underbrace{\int_0^\infty \left[ \frac{W^{*^2}}{2} \right] r^*dr^*}_{\text{III}},
\label{eq:drag_coeff_balance}
\end{multline}

\noindent where $\rho$ is the working fluid density, $A_{\mbox{disc}} = \pi D^2/4$ the surface area of the actuator disc, $F_D$ its aerodynamic drag and $\Delta U = U_\infty - U$ the velocity defect. Eq. \ref{eq:drag_coeff_balance} relates the drag coefficient $C_D$ to the mean momentum deficit flow rate (term $I$), the turbulence anisotropy (term $II$) and the mean kinetic energy of swirl (term $III$). Our measurements emphasise that term $I$ predominates Eq. \ref{eq:drag_coeff_balance} which therefore reduces to

\begin{equation}
C_D \approx 16\int_0^\infty U^\star \Delta U^\star  r^\star dr^\star.
\label{eq:CD_FW}
\end{equation}

\noindent In the same manner, using Eq. \ref{eq:phi_momentum}, an angular momentum budget applied on the control volume $\mathcal{V}$ yields

\begin{equation}
G_0^\star = \frac{2 G_0}{\rho U_\infty^2 A_{\mbox{disc}} D} = 16 \int_0^\infty ( U^\star W^\star + \overline{u'w'}^\star) r^{\star^2}dr^\star,
\label{eq:G0_dimensionless}
\end{equation}

\noindent which expresses that the integrated angular momentum remains constant along the streamwise direction and is equal to its source value. Here again, our measurements show that the transport of angular momentum by the Reynolds shear stress (i.e. $\overline{u'w'}^\star$) is marginal compared to that transported by the mean swirling motion (i.e. $U^\star W^\star$). Eq. \ref{eq:G0_dimensionless} thus reduces to

\begin{equation}
G_0^\star \approx 16 \int_0^\infty U^\star W^\star r^{\star^2} dr^\star.
\label{eq:G0_FW}
\end{equation}
           
\noindent The conservation laws (\ref{eq:CD_FW}) and (\ref{eq:G0_FW}) govern the evolution of the mean swirling wake as it develops in the streamwise direction and are at the basis of the self-preservation analysis in the following part. 
    
\subsection{Similarity analysis}

\noindent The similarity analysis consists in seeking self-similar solutions for the flow properties which have to satisfy the conservation laws (\ref{eq:CD_FW}) and (\ref{eq:G0_FW}). These conservation laws act as the "\textit{similarity constraints}", as stated by \cite{George89}. To this end, we assume that any flow variable can be expressed as $\bullet(x,r) = \bullet_s(x) f(\zeta)$ where $\bullet_s$ represents the typical amplitude of the variable, $f$ is a universal self-similar function and $\zeta = r/\delta(x)$ with $\delta(x)$ a characteristic length of the flow. An important remark regarding the choice of $\delta$ has to be made here. Since the evolution of classical axisymmetric wake is solely governed by the linear momentum conservation law (\ref{eq:CD_FW}), it is natural to choose a characteristic length directly connected to the velocity deficit $\Delta U$. For that reason, $\delta$ is most often assimilated to the wake half-width $\delta_{1/2}$, which is by definition $\Delta U(x, r=\delta_{1/2}) = U_s(x)/2$ where $U_s(x)$ is the maximum velocity deficit at the streamwise location $x$ \citep{Pope2000}. Unlike classical axisymmetric wakes, the swirling wake is featured by an additional constraint in the form of the conservation of mean angular momentum (\ref{eq:G0_FW}). This enables the swirling wake to scale with another characteristic length such as $\delta_{swirl}$ which features the swirling motion such that $W(x,r=\delta_{swirl}) = W_s(x)$ where $W_s(x)$ is the swirl amplitude at the streamwise location $x$. In other words, the way the swirling wake develops will depend on the competition between the linear and angular momentum. In the following, we assume that the wake in the vicinity of the actuator disc is a region of the flow governed by the swirling motion.

Substituting self-similar forms in the conservation laws (\ref{eq:CD_FW}) and (\ref{eq:G0_FW}) and assuming that the porosity of the actuator disc is high enough so that $U^\star \approx 1$, it comes

\begin{align}
\label{eq:SimCst_CD}
C_D \sim \left[ U_s^\star \delta^{\star^2} \right] \int_0^\infty \zeta g(\zeta) d\zeta, \\
\label{eq:SimCst_G0}
G_0^\star \sim \left[ W_s^\star \delta^{\star^3} \right] \int_0^\infty \zeta^2 h(\zeta) d\zeta,
\end{align}

\noindent where $g$ and $h$ are similarity functions. Given that the linear and angular momentum remain constant and equal to their source value, the products in square brackets in the previous equations are also constant, which implies that

\begin{align}
\label{eq:SimLaw_CD}
U_s^\star \sim \delta^{\star^{-2}}, \\
\label{eq:SimLaw_G0}
W_s^\star \sim \delta^{\star^{-3}}.
\end{align}

\noindent Using (\ref{eq:SimLaw_CD}) and (\ref{eq:SimLaw_G0}) yields $U_s \sim W_s^{2/3}$, a relationship that is well supported by the data reported in \citet{Wosnik2013}. A final step is then necessary to close the system. To this end, the transport equation of the turbulent kinetic energy is used, providing an additional constraint relating the expansion rate of the characteristic length scale $\delta$ to the turbulent dissipation rate $\epsilon$ and reads \citep{George89}

\begin{equation}
\frac{d\delta}{dx} \sim \frac{\epsilon \delta}{U_s^2 U_\infty}.
\label{eq:deltarate}
\end{equation}

\noindent A general expression for $\epsilon$ was proposed by \citet{vassilicos2015dissipation} and reads

\begin{equation}
\epsilon = C_\epsilon \frac{U_s^3}{\delta},
\label{eq:dissipation}
\end{equation}

\noindent with $C_\epsilon \sim Re_D^m / Re_\ell^n$, where $Re_D = U_\infty D / \nu$ and $Re_\ell = U_s \delta / \nu$ represent global and local Reynolds numbers, respectively. The nature of the turbulence is then controlled by the exponents $m$ and $n$. Following \citet{dairay_non-equilibrium_2015}, the classical equilibrium turbulence (in the Kolmogorov sense) is retrieved by imposing $m=n=0$, which yields $C_\epsilon = \mbox{const}$, while non-equilibrium turbulence corresponds to the condition $m=n=1$. Coupling Eqs. \ref{eq:deltarate} and \ref{eq:dissipation} yields

\begin{equation}
\frac{d\delta^\star}{dx^\star} \sim Re_D^{m-n} \frac{U_s^{\star^{1-n}}}{\delta^{\star^{n}}}.
\label{eq:deltarate_ENE}
\end{equation}

\noindent With $n=m=0$, Eq. \ref{eq:deltarate_ENE} becomes $d\delta^\star/dx^\star \sim U_s^\star$ meaning that the expansion rate scales linearly with the velocity deficit. Substituting \ref{eq:SimLaw_CD} into \ref{eq:deltarate_ENE} yields the classical so-called \citet{Townsend1976} \& \citet{George89} scalings 

\begin{equation}
\delta^\star \sim (x^\star - x_0^\star)^{1/3} \quad \mbox{and} \quad U_s^\star \sim (x^\star - x_0^\star)^{-2/3},
\end{equation}

\noindent where $x_0$ stands for a virtual origin. Accounting for \ref{eq:SimLaw_G0}, it immediately follows that

\begin{equation}
W_s^\star \sim (x^\star - x_0^\star)^{-1}.
\end{equation}

\noindent This scaling law, which dictates the decay of the swirl amplitude, was initially established by \citet{Wosnik2013}. However, the authors noticed that this prediction was not well supported by their experimental data. They argued that this departure from the predicted law might have been related to the presence of tip vortices. Recently, \citet{holmes2022impact}, who studied the axisymmetric swirling wake of a rotating porous disc at different rotating speeds, claimed that the equilibrium scaling laws predicted the swirl decay rate only after an "\textit{initial adjustment region}". In other words, \citet{Wosnik2013} as well as \citet{holmes2022impact} evidenced that equilibrium similarity fell short in accurately predicting swirl decay in the intermediate wake region. In fact, this region of the wake is characterised by strong non-homogeneity and anisotropy, where equilibrium is highly unlikely \citep{vassilicos2015dissipation}.

\noindent Accordingly, considering now the non-equilibrium framework, i.e. $n=m=1$, Eq. (\ref{eq:deltarate_ENE}) becomes $d\delta^\star/dx^\star \sim \delta^{\star^{-1}}$, which implies

\begin{equation}
\delta^\star \sim (x^\star - x_0^\star)^{1/2}.
\label{eq:deltascaling_NE}
\end{equation}

\noindent Note that neither of the similarity constraints (\ref{eq:SimLaw_CD}) nor (\ref{eq:SimLaw_G0}) were invoked to obtain Eq. (\ref{eq:deltascaling_NE}). This means that, with a non-equilibrium approach, the nature of the length scale $\delta$ is not implicitly attributed either to the velocity deficit or the swirling motion. Instead, $\delta$ is directly related to the mechanism which sets the level of dissipation in the turbulent wake. Injecting (\ref{eq:deltascaling_NE}) into (\ref{eq:SimLaw_CD}) and (\ref{eq:SimLaw_G0}) yields

\begin{equation}
U_s^\star \sim (x^\star - x_0^\star)^{-1}, 
\end{equation}

\noindent and

\begin{equation}
    W_s^\star \sim (x^\star - x_0^\star)^{-3/2}.
    \label{eq:Ws_novel_scaling}
\end{equation}

\noindent This novel non-equilibrium scaling law (\ref{eq:Ws_novel_scaling}) predicts a faster decay of swirl amplitude when compared to its equilibrium counterpart. Naturally, this theoretical law needs to be confronted with data for validation. In this work, an experimental wind tunnel approach was privileged. The following section will therefore provide a comprehensive description of the experimental set-up used to generate and characterise the swirling wake of a modified actuator disc.

\section{Experimental set-up}\label{sec:experimental_setup}

\subsection{Design of the porous discs with passive swirl generation}\label{subsec:passive_swirl_gen}

\noindent The actuator discs used in this study have been designed based on the work of \citet{helvig_2021} who investigated the wake generated by porous discs featuring various porosity patterns. Here, following the nomenclature proposed by \citet{helvig_2021}, a scaled-up version of their Non-uniform Holes Disc with 35\% solidity (referred to as NHD35 in \citet{helvig_2021}) has been selected as a reference. The main geometrical parameters and dimensions of the porous discs are shown in Fig. \ref{fig:PD_GP}. The porous discs have a diameter and a thickness of $D$ = 100 mm and $e$ = 5 mm, respectively. Each porous disc has a central solid disc of $25$ mm in diameter with a $6$ mm hole in the middle. From this solid disc, 16 trapezoidal blades make up the body of the disc along with an inner rim of $61$ mm in diameter and $1.5$ mm in width. The outer rim of the porous disc is $2$ mm wide. Following \cite{Camp2016} and \cite{helvig_2021}, the number of blades and their size are the main factors which can be adjusted to obtain the desired porosity, noted $\beta$, defined as: 

        \begin{figure}
            \centering
            \begin{subfigure}{0.55\textwidth}
                \includegraphics[width=\textwidth]{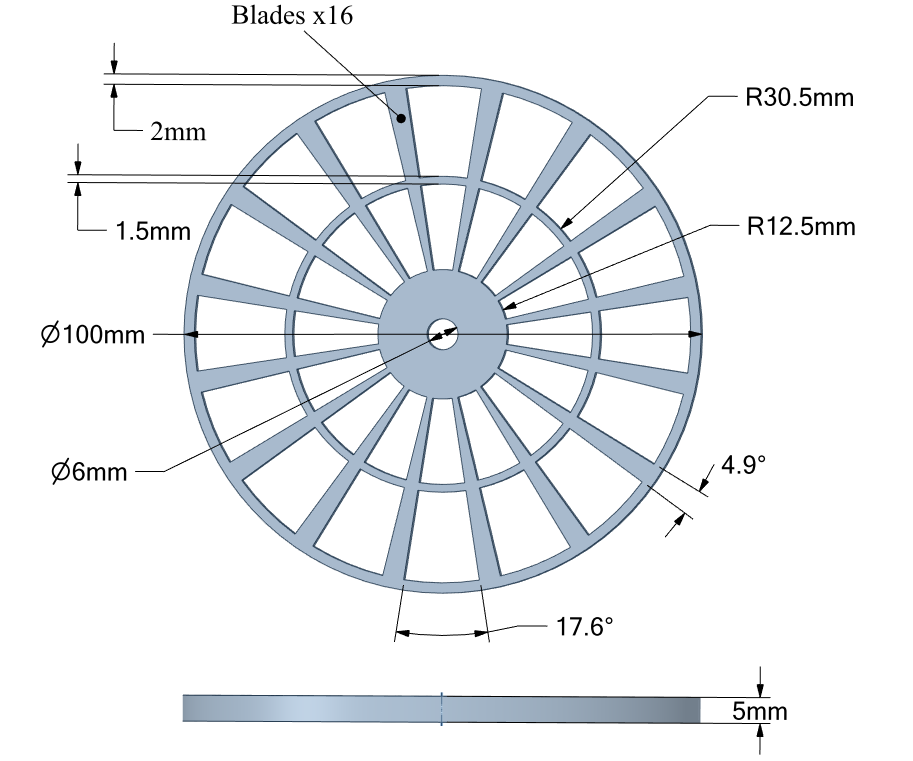}
                \subcaption{}
            \end{subfigure}
            \begin{subfigure}{0.28\textwidth}
                \includegraphics[width=\textwidth]{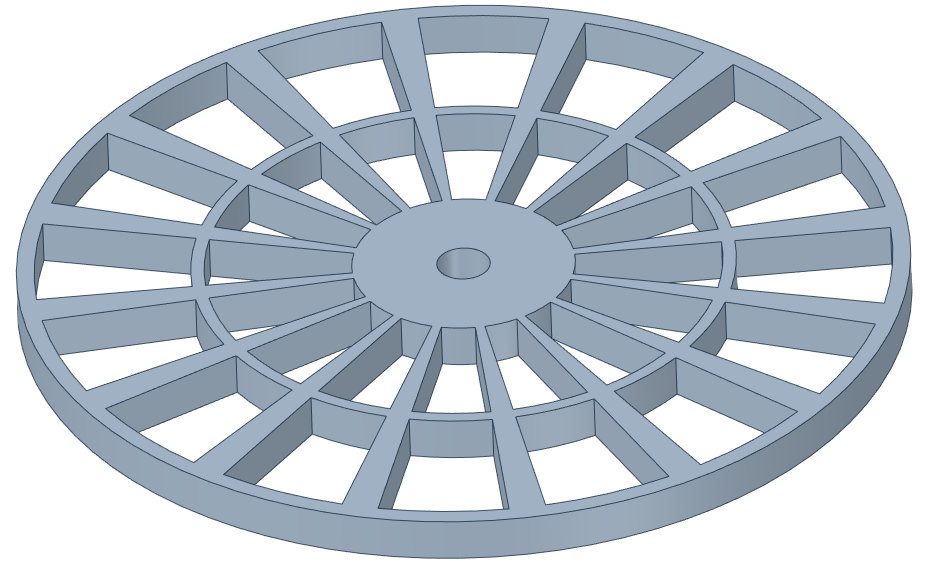}
                \vspace{2cm}
                \subcaption{}
                
            \end{subfigure}
            
            \caption{NHD35 porous disc (a) global parameters (front and side view) and (b) isometric view}
            \label{fig:PD_GP}
        \end{figure}

        \begin{equation}
            \beta = \frac{A_h}{A_{\mbox{disc}}},
            \label{eq:porosity_def}
        \end{equation}

\noindent where $A_{h}$ is the empty area of the trapezoidal holes. While the local porosity varies along the radial direction, the global porosity of the disc is $\beta=65\%$, matching exactly the NHD35 disc studied in \cite{helvig_2021}. The porous discs were designed with a commercial 3D design software (ANSYS\textsuperscript{TM} SpaceClaim) and printed in Polylactic acid (PLA) using a Cura\textsuperscript{TM} Ultimaker 3 Extended printer.
        
         \begin{figure}
            \centering
            \begin{subfigure}{0.4\textwidth}
               \includegraphics[width=\textwidth]{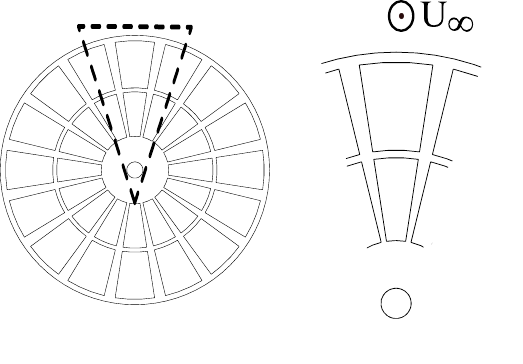} 
               \subcaption{}
            \end{subfigure}
            \begin{subfigure}{0.4\textwidth}
               \includegraphics[width=\textwidth]{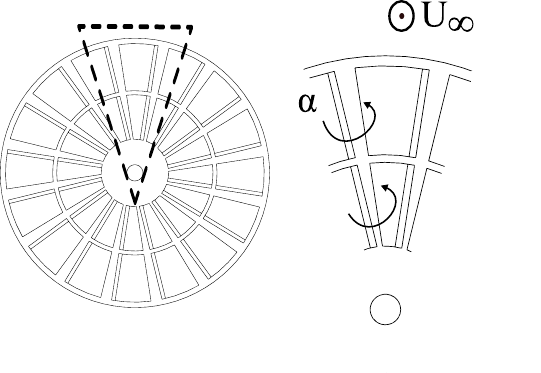} 
               \subcaption{}
            \end{subfigure}

            \vspace{0.3cm}
            
            \centering
            \begin{subfigure}{0.6\textwidth}
               \includegraphics[width=\textwidth]{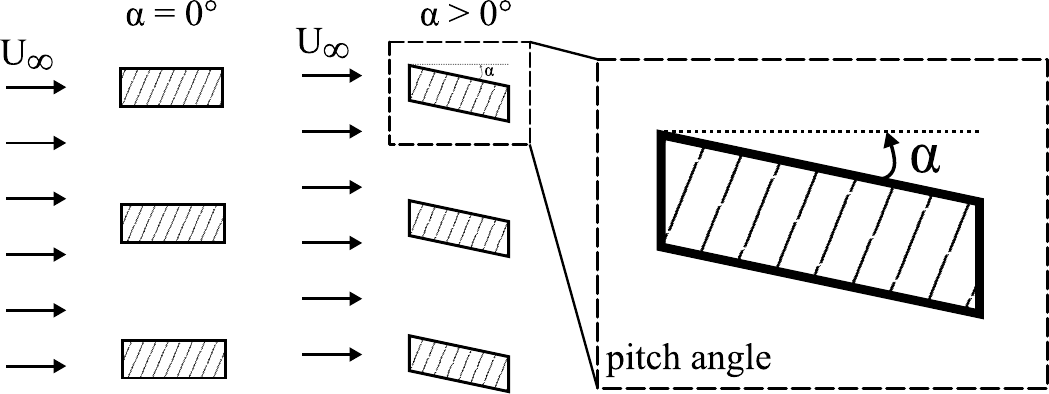} 
               \vspace{0.1cm}
               \subcaption{}
            \end{subfigure}

            \caption{Modification of the porous discs to generate a swirling wake showing a frontal view of (a) the $\alpha = 0  \degree$ reference case, (b) of a porous disc with pitched blades ($\alpha >0 \degree$) and (c) an unfolded view of the modifications. Black dashed lines: zoomed-in area.}
            \label{fig:PD_swirl_gen}
        \end{figure}

\begin{figure}
	\centering
         \includegraphics[width=0.9\textwidth]{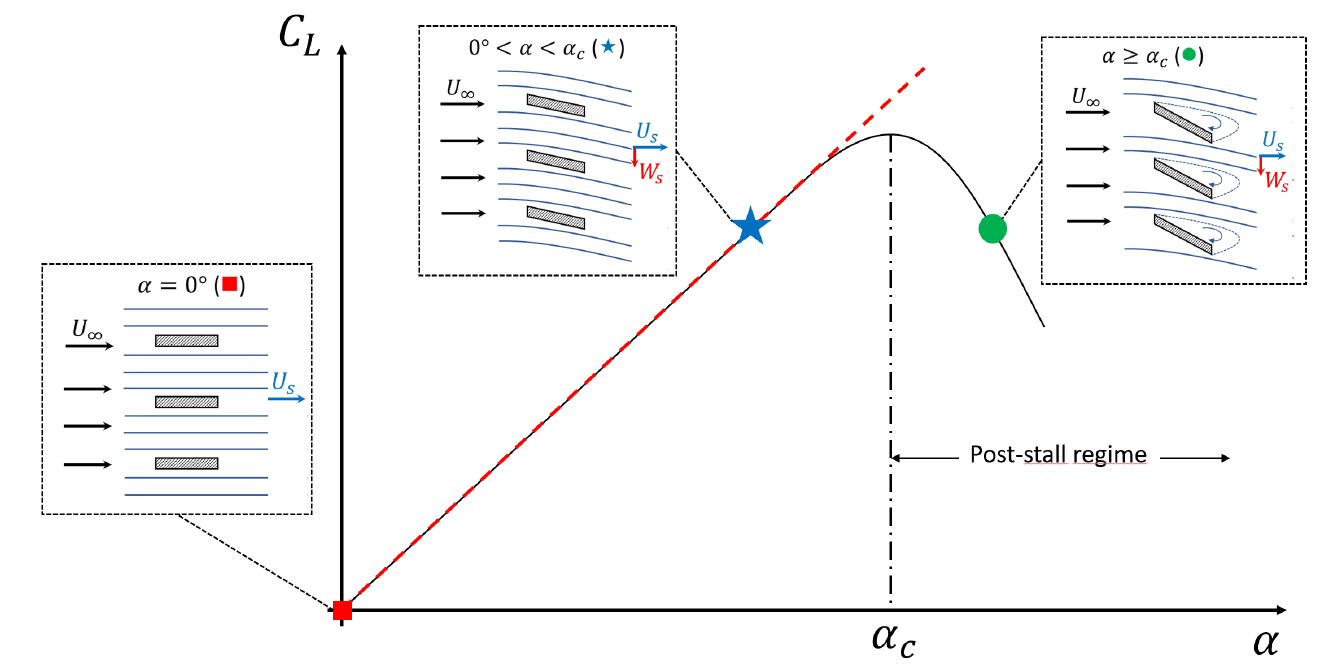}
	\caption{Schematic of the influence of the pitch angle $\alpha$ on the lift coefficient experienced by the blades of the porous disc to passively add swirl. The thin airfoil theory prediction is symbolized by the red dashed line. The vertical black dot-dashed line represents the critical angle at which stall occurs. Unfolded views are provided to illustrate the flow path through the actuator disc with different scenarios: (\textcolor{red}{$\blacksquare$}) blades with no pitch, (\textcolor{blue}{$\bigstar$}) blades with a pitch angle in the pre-stall regime and (\textcolor{green}{$\CIRCLE$}) with a pitch angle in the post-stall regime.}
	\label{fig:CLAlphaDiagram}
\end{figure}

\noindent In contrast to the approach taken by \citet{holmes2022impact}, we chose to introduce the swirling motion in a passive manner by slightly modifying the disc's geometry. This methodology is illustrated schematically in Fig. \ref{fig:PD_swirl_gen}. It consists in pitching the trapezoidal blades that come out from the centre of the disc of an angle $\alpha>0\degree$. The main interest of this approach is to keep the design stage simple and inexpensive. However, this method lacks pre-defined control over the level of injected swirl and must be therefore estimated \emph{a posteriori}. Nevertheless, some predictions for the swirl amplitude as a function of $\alpha$ can be made by assimilating each individual blade to a thin flat plate of infinite span for which the thin airfoil theory \citep{anderson2011fundamentals} predicts a lift coefficient $C_L = 2\pi \alpha$, with $\alpha$ expressed in radians, as depicted in Fig. \ref{fig:CLAlphaDiagram}. Besides, according to the Kutta-Joukowski theorem, for non-stalled blades, the lift coefficient per unit span can be expressed as a function of the circulation $\Gamma$ such as $C_L = 2\Gamma^\star/e^\star$, where $e^\star$ is assimilated to the dimensionless blade chord. Since the inclination of the blades is what induces the swirling motion, the circulation per unit span is intimately related to the swirling velocity such as $\Gamma \sim W_s \ell$ where, by definition $\ell$ is the typical extent of the contour integral on which the circulation is computed. It comes naturally that $\ell = \pi D/n$, where $n$ is the number of blades of the porous disc, which leads to
        
        \begin{equation}
            C_L \sim \left(\frac{2\pi}{n}\right) \frac{W_s^\star}{e^\star}.
            \label{eq:CL_f_W}
        \end{equation}

\noindent Therefore, the amplitude of the swirling motion featured by $W_s^\star$ scales linearly with the lift coefficient per unit span $C_L$ according to potential flow theory. Replacing the expression for the lift coefficient derived from thin airfoil theory in Eq. (\ref{eq:CL_f_W}) yields
    
        \begin{equation}
            W_s^\star \ \sim \ n e^\star \alpha = 0.8 \alpha,
            \label{eq:Ws_link_alpha}
        \end{equation}

\noindent which predicts a linear increase of the swirl magnitude with respect to the pitch angle $\alpha$. The slope of Eq. (\ref{eq:Ws_link_alpha}) sets the theoretical upper limit for the generation of swirl until stall appears. Due to finite size blade effects, non constant aspect ratio of the blades and their surface finish, a lower slope is expected. As shown in Fig. \ref{fig:CLAlphaDiagram}, there is a critical pitch angle $\alpha_c$ beyond which stall appears. This will result in a decrease in lift, consequently reducing the swirl intensity according to (\ref{eq:CL_f_W}). The value of $\alpha_c$ depends on several parameters (e.g., Reynolds number, surface roughness, aspect ratio ...), making it difficult to accurately predict. However, based on experimental data reported in the literature, a discernible range can be defined for flat plates with comparable average aspect ratios, as highlighted by \cite{nakayama1988visualized,mohebi2017turbulence}, this range is $\alpha_c \in [14\degree - 20\degree]$. To tackle this issue, a parametric study has been conducted by varying the pitch angle $\alpha$ within the range $[5\degree - 30\degree]$. Moreover, the appearance of massive flow separation over the blades will diminish the effective porosity of the actuator disc, leading to an increase in its drag. To evaluate this effect, supplementary discs with thicker, but not pitched, blades were manufactured. For those discs, the blade thickness was chosen such that their frontal area matches the projected area of an original blade at a specific pitch. These low-porosity discs have been manufactured to be compared to pitched discs at $\alpha = 15\degree$ and $\alpha = 25\degree$. All the 3D models of the porous discs used in this work are available in the supplementary material.

\subsection{Wind tunnel and test rig implementation}

\noindent The experiments were conducted in the S2 subsonic, Eiffel type, open-circuit wind tunnel of the PRISME Laboratory at the University of Orléans. The test section is 2 m long with a cross-sectional area of 0.50$\times$0.50 m$^2$ with walls entirely made of Plexiglas to allow for optical access. The free-stream velocity $U_\infty$ can reach 50 m.s$^{-1}$, while the background turbulence intensity remains lower than $0.35\%$.

\begin{figure}

\begin{subfigure}{0.18\textwidth}
    \centerline{\includegraphics[width=\textwidth]{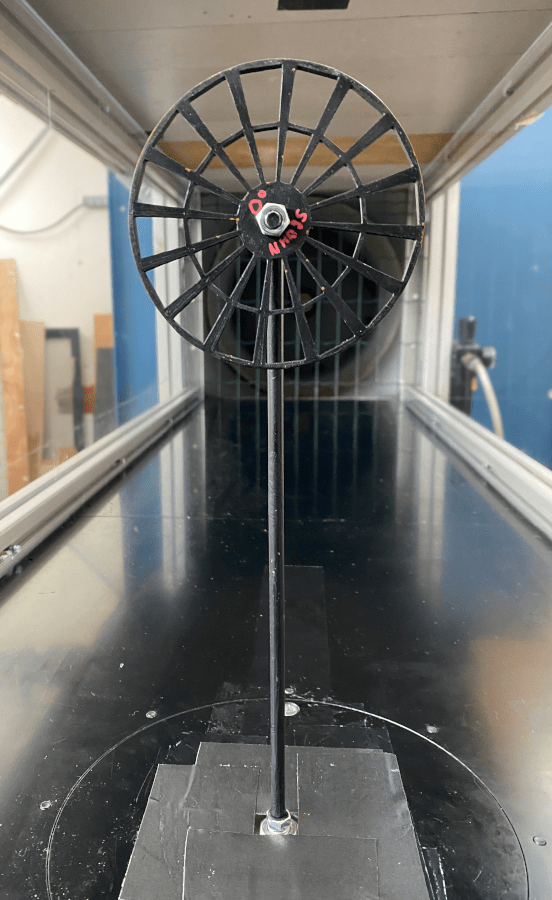}}
    \caption{}
\end{subfigure}
\begin{subfigure}{0.8\textwidth}
    \centerline{\includegraphics[width=\textwidth]{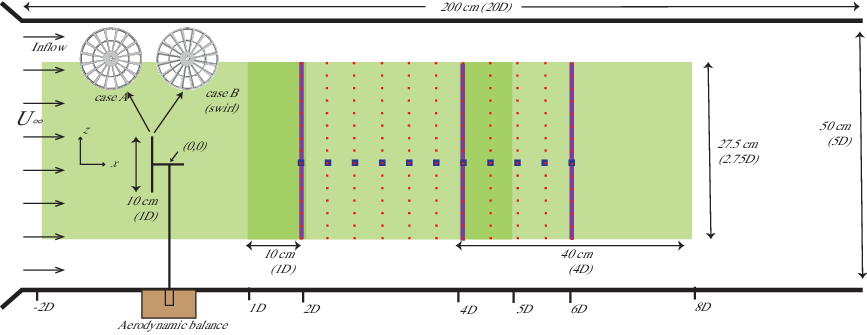}}
    \caption{}
\end{subfigure}
  
  \caption{(a) Front view of the test rig and (b) schematic of the experimental set-up (not to scale). Green boxes: PIV fields of view, magenta solid lines: maps in $y$-$z$ planes, dotted red lines: vertical ($z$) profiles, blue squares: horizontal ($y$) profiles (3CHWA).}
\label{fig:exp_set}
\end{figure}

\noindent The implementation of the porous discs in the wind tunnel is depicted in Fig. \ref{fig:exp_set}. The discs are fixed on a T-shaped aluminium cylindrical rod with a diameter $d$ = 6 mm representing the wind turbine nacelle and tower. The hub can be assimilated to the central solid disc where the disc is fixed to the rod (Fig. \ref{fig:PD_GP}). With this set-up, the porous discs were interchangeable without the need to remove the rod, ensuring that our data would not be affected by an eventual misalignment between experiments. The wind tunnel blockage ratio is approximately $2\%$ (rod included) ensuring a relatively undisturbed wake. Accordingly, no blockage correction was applied. The test rig is mounted to an aerodynamic balance placed below the wind tunnel floor. The origin of the coordinates is located at the crossroads between the mast and the nacelle rod. The $x$-axis corresponds to the streamwise direction, the $y$-axis to the spanwise direction and the $z$-axis to the vertical direction.

\subsection{Metrology and methodologies}

\noindent The measurement tools used in this work were selected to assess the conservation laws and similarity analysis established in \S\ref{sec:similarity}. Accordingly, the drag force experienced by the porous disc has to be measured to obtain direct values of $C_D$. Besides, three-components velocity measurements are needed to fully characterise the wake evolution.
             
\subsubsection{Drag force measurements}

\noindent In order to determine the drag coefficient $C_D$ of each porous disc, an ATI\textsuperscript{TM} Mini40-E balance was used. Placed directly beneath the wind tunnel floor, this balance is capable of measuring force and torque components in all three directions ($x,y,z$). Since the entire test rig is attached to the Mini40-E attach point, the balance provides measurements of the test rig's total drag $F_x^{TR}$. To isolate the drag experienced by the porous disc from that generated by the mast (excluded from our definition of $C_D$), we followed the method proposed in \cite{helvig_2021}. It consists in performing drag measurements of the T-shaped rod alone $F_x^{rod}$ and subtracting them from the test rig's total drag, giving thereby an estimation of the porous disc drag following $F_D = F_x^{TR} - F_x^{rod}$. The sampling frequency was set to 1 kHz and the measurement duration was $180$ s. A moving average on 100 samples was used to filter out the vibrations, resulting in an effective sampling frequency of $10$ Hz. The uncertainties were estimated using the calibration errors ($\pm 1.5\%$ of the measured load) and the statistical errors. The statistical errors were determined using the standard deviation of the force signal. The uncertainties amounted to $\epsilon_{C_D}=0.03$ on average. It was found that beyond $Re_D \approx 10^5$, $C_D$ becomes Reynolds number independent. Therefore, in the remainder of this paper, we report only results obtained at $Re_D = 1.3 \times 10^5$, which corresponds to a free-stream velocity of $U_\infty = 20$ m.s$^{-1}$.

\subsubsection{Three components hot-wire anemometry}

\begin{figure}
  \centerline{\includegraphics[width=0.9\textwidth]{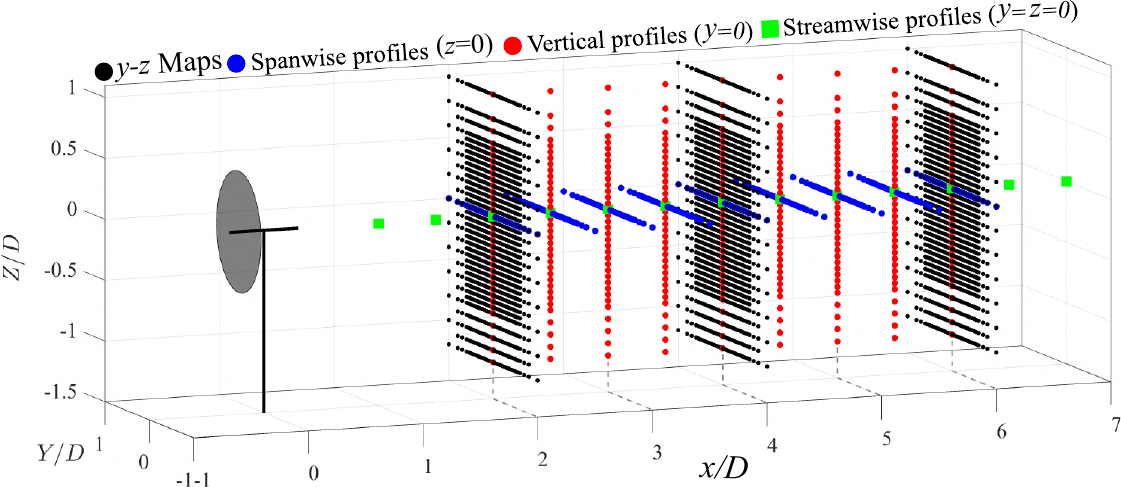}}
  \caption{3CHWA measurement points. Grey disc: porous disc position, solid black line: cylindrical mast.}
\label{fig:hwa_detail}
\end{figure}

\noindent To measure the swirling wake, 3-components hot-wire anemometry (3CHWA) measurements were conducted using a Dantec Dynamics\textsuperscript{TM} Streamline constant temperature anemometry (CTA) system with a gold-plated tungsten tri-axial wire probe (Dantec Dynamics\textsuperscript{TM} 55P91 probe). The probe has three gold-plated tungsten wires having a diameter of 5 $\mu$m, an individual wire sensing length of 1.2 mm and a total sensing length of 3.2 mm, which is of order of the Taylor micro-scale $\lambda \in [2 ; 4] $ mm \citep{sreenivasan1983zero,mora2019energy}. The probe was subjected to a directional calibration before each campaign and to a velocity calibration between measurements. This specific probe has a maximum yaw angle range of $[-30\degree ; +30\degree]$ beyond which velocity measurements are erroneous. The HWA measurements were used to estimate the characteristic scales of the turbulent flow which are reported in table \ref{tab:turbulence_stats}. The estimated scales are: $\mathcal{L}_{int}$ the integral length scale, $\lambda$ the Taylor micro-scale, $\eta$ the Kolmogorov length scale and $\mathcal{T}_{int}=\mathcal{L}_{int}/U$ the integral time scale.

         \begin{table}
                \centering
                 \begin{tabular}{ *8l }    \toprule
                $\alpha (\degree)$ & 0 & 5 & 10 & 15 & 20 & 25 & 30 \\\midrule
                $\mathcal{L}_{int}$ (mm) & 16 & 15 & 16 & 17 & 23 & 28 & 30\\[10pt]
                $\lambda$ (mm) & 3 & 3 & 3 & 3 & 2 & 2 & 3\\[10pt]
                $\eta$ ($\mu$m) & 250 & 230 & 262 & 220 & 230 & 250 & 290\\[10pt]
                $\mathcal{T}_{int}$ (ms) & 1.4 & 1.3 & 1.4 & 1.6 & 1.9 & 3.2 & 2.5\\\bottomrule
                 \hline
                 
                \end{tabular}
                
                \caption{Typical turbulent length and time scales calculated in the wake of the different porous discs at $x^\star=6, y^\star = z^\star = 0$.}
                \label{tab:turbulence_stats}
        \end{table}

\noindent As shown in Fig. \ref{fig:hwa_detail}, cross-sectional planes were measured at three streamwise positions: $x=2D$, $x4D$ and at $x=6D$. The number of measurements points is $N_p=1130$. The spatial resolution is $\Delta y_{wake} = \Delta z_{wake} = 5$ mm ($\approx 2\lambda$) inside the wake and 10-20 mm ($\approx \mathcal{L}_{int}$) in the free stream (figure \ref{fig:hwa_detail}). Spanwise and vertical profiles were measured between $X=2D$ and $X=6D$ with a streamwise step of $\Delta_x=0.5D$. The sampling frequency was set to $40$ kHz for an acquisition time of $2.5$ s ($\sim 10^3 \mathcal{T}_{int}$) at each location. Streamwise profiles ($y^\star=z^\star=0$) were also measured from $x^\star=1$ to $x^\star=7$. During the HWA measurements, the main uncertainty sources came from the inherent changes in the experimental conditions (temperature, humidity, pressure) and from the calibration of the hot-wire probe \citep{huffman1980calibration,bruun1996hot}. The total uncertainty in the free-stream was below 0.5\% and around 0.8-2\% in the wake depending on the position of the probe with respect to the test rig.  The temperature of the room was monitored throughout each experiment and was shown not to exceed a variation of more than $1^\circ C$ for all cases. The HWA uncertainties were estimated between experiments using the calibration unit and corroborated by redundant measurements. Following Eq. (\ref{eq:CD_FW}) and to better assess the effect of swirl on the flow topology, the streamwise velocity component had to be measured with a better spatial resolution and accuracy that what is achievable with HWA. Therefore, Particle Image Velocimetry (PIV) measurements were performed in the streamwise plane.

\subsubsection{Particle image velocimetry over multiple fields of view}

\noindent Planar particle image velocimetry (PIV2D2C) was used to characterise the generated wakes in a $x-z$ plane (figure \ref{fig:exp_set}). In the streamwise direction, the region of interest is located between $x=-2D$ and $x=8D$. This region is decomposed into 3 separate fields of view (FoV) with an overlapping area of 25\% ($1D$). The dimensions of the individual FoVs are ($x \times z$) $4.2D\times 2.6D$ and the full region reaches $10D \times 2.6D$. The total region was obtained by merging the mean velocity fields and standard deviations using a similar approach to that proposed in \citet{Li2021}. The image sets were captured using an 11 megapixels LaVision\textsuperscript{TM} LX-11M CCD camera mounted on an optical rail parallel to the test section. A ZEISS\textsuperscript{TM} camera lens was used with focal length $f_0 = 85$ mm and was set at an aperture of $f_0/4$. The flow was illuminated using a double pulse Nd:YAG (532 nm) laser system generating a laser sheet of 1.5 mm in thickness. The flow was seeded with olive oil droplets from an aerosol generator and had an average diameter of $d_p\approx 2-3$ $\mu$m. Olive oil droplets were chosen as tracers since they are non-reactive, non-toxic, scatter light appropriately and are sufficiently small in order to faithfully represent the fluid motion. To prove this last important point, the average Stokes number $St$ was calculated using all scales of motion and showed that $St \in [1 ; 40] \times 10^{-3} \ll 1$. The generated tracer particles will therefore follow all scales of motions reliably \citep{kallio1992interaction,vincent2007aerosol}.

\noindent For each FoV, 2600 image pairs were recorded at a time interval of $dt = 55$ $\mu$s between snapshots. The sampling frequency was set at $f_{PIV}=2.1$ Hz rate which corresponds to a total acquisition time of $T_{PIV} = 24 $ minutes. The laser pulses and the frame recordings were synchronised using an external LaVision\textsuperscript{TM} Programmable Timing Unit (PTU). The snapshots were analysed using a commercial PIV software (Davis 10.2, LaVision\textsuperscript{TM}). A multi-pass cross-correlation method was applied using an initial interrogation window (IW) size of $64\times64$ pixels and a final IW of $32\times32$ with 50\% overlap \citep{Raffel2007}. A gaussian filter was used for sub-pixel interpolation and a median filter was applied in order to remove eventual spurious vectors. 

 Each FoV required an individual calibration. The calibration was performed using a custom calibration plate with uniformly spaced dots. The plate is 540 mm long and 420 mm tall and has a total of 560 dots ($28 \times 20$) which are 4 mm in diameter and 19.45 mm apart. The calibration plate was placed in the mid-span plane of the test section next to the test rig. The resulting calibrations allowed each field of view to have a magnification factor (pixels to mm) and to correct optical aberrations \citep{Raffel2007}. The magnification factor had a constant value of $SF=9.61$ pixels/mm for each calibration. The PIV algorithm resulted in a resolution of $\Delta_x^{\text{PIV}} = \Delta_z^{\text{PIV}} = 1.66 $ mm which is of the order of the Taylor micro-scale $\lambda$ and 10 times smaller than the integral length scale $\mathcal{L}_{int}$ (see table \ref{tab:turbulence_stats}).

Uncertainties for the PIV measurements were calculated using correlation statistics, a method presented in \cite{Wieneke2015PIV} and applied here. This method estimates PIV uncertainties based on a pixel-wise statistical analysis that quantifies the contribution of each pixel to the shape of the correlation peak. The error quantification method resulted in a displacement uncertainty of $\epsilon_d \in [0.03 ; 0.08]$ px which fall within the order of magnitude of the 0.06 px value recommended in \cite{Raffel2007}. As a side note, the error shoots up to values of $\epsilon_d \approx 0.15-0.20$ px very close to the test rig which is expected since strong 3D effects are present and increase the noise due to out-of-plane motion. Moreover, the swirling cases will increase this out-of-plane motion by definition. This error was quantified and corresponds to an instantaneous velocity uncertainty of 0.8\% and a mean uncertainty below 0.5\% in the wake of the porous discs. It was verified that the flow statistics converged properly beyond 2000 snapshots.

\section{Modified porous disc aerodynamics}\label{sec:aero}

\noindent This section is devoted to a parametric study of how the pitch angle $\alpha$ affects the aerodynamics of the porous discs presented in the previous section, with reference to the unmodified porous disc for which $\alpha=0\degree$. First, attention is focused on the evolution of the swirl characteristics with respect to $\alpha$, from which two distinct operating regimes are highlighted. Then, drag and swirl number induced by the modified discs are analysed in detail with the aim to distinguish the effects of porosity and swirl.

\subsection{Characteristics of the swirling motion}

  \begin{figure}
               \centering
                    \begin{subfigure}{0.282\textwidth}
                    \centerline{\includegraphics[width=\textwidth]{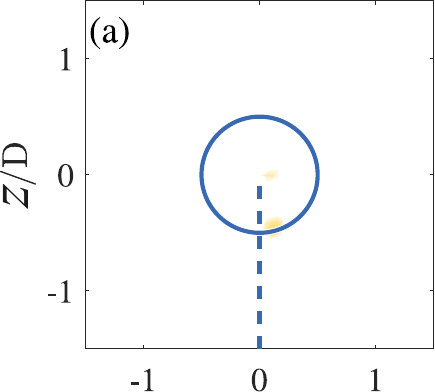}}
                    \end{subfigure}
                     \hspace{0.25\textwidth}
                    \begin{subfigure}{0.358\textwidth}
                      \centerline{\includegraphics[width=\textwidth]{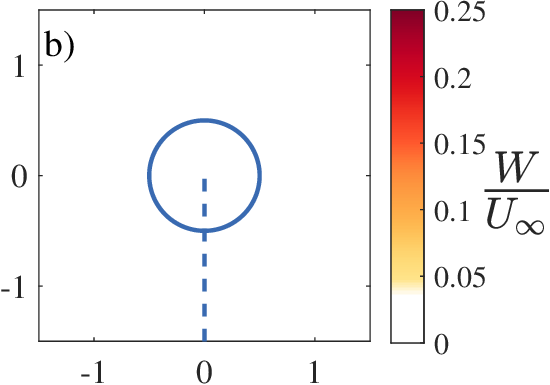}}
                     \end{subfigure}
    
                      \centering
                      \begin{subfigure}{0.28\textwidth}
                      \centerline{\includegraphics[width=\textwidth]{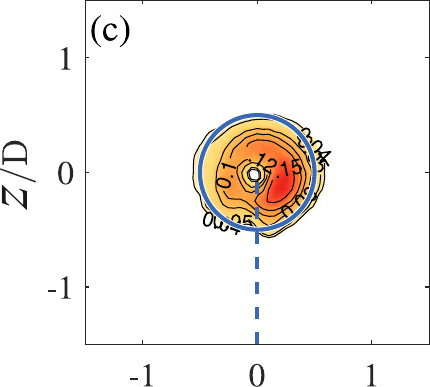}}
                      \end{subfigure}
                      \begin{subfigure}{0.25\textwidth}
                      \centerline{\includegraphics[width=\textwidth]{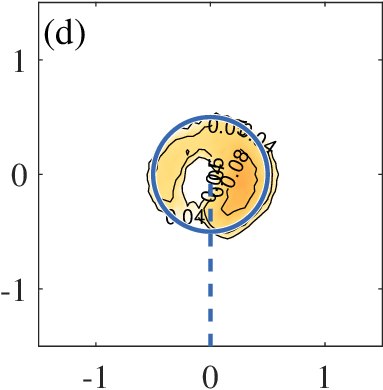}}
                      \end{subfigure}
                      \begin{subfigure}{0.355\textwidth}
                      \centerline{\includegraphics[width=\textwidth]{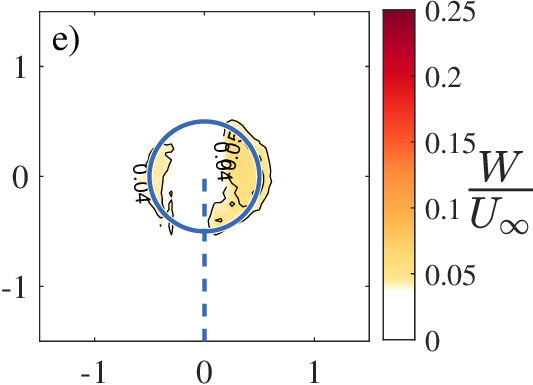}}
                      \end{subfigure}
    
                      \centering
                      \begin{subfigure}{0.279\textwidth}
                      \centerline{\includegraphics[width=\textwidth]{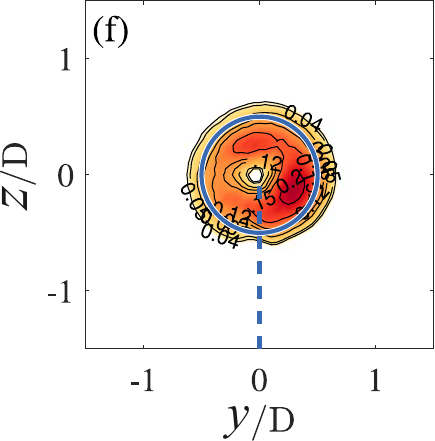}}
                      \end{subfigure}
                      \begin{subfigure}{0.248\textwidth}
                      \centerline{\includegraphics[width=\textwidth]{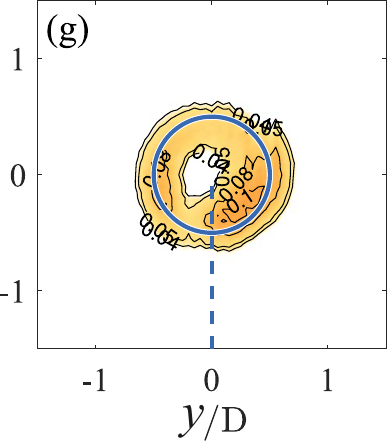}}
                      \end{subfigure}
                      \begin{subfigure}{0.362\textwidth}
                      \centerline{\includegraphics[width=\textwidth]{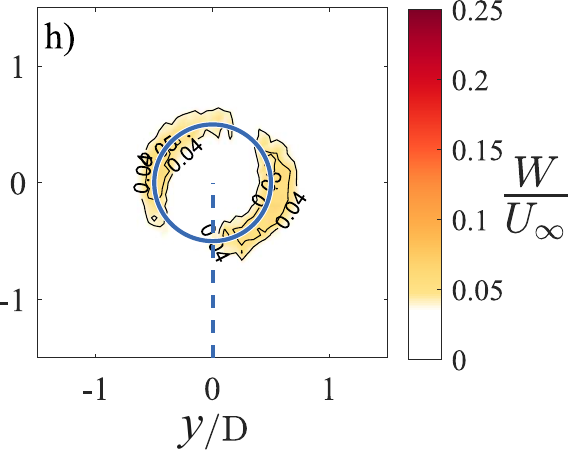}}
                      \end{subfigure}
      
                    \caption{Normalised swirling velocity $W^\star$ spanwise maps for the (a, b) $\alpha=0\degree $ case, (c, d, e) the $\alpha=15\degree $ case and (f, g, h) the $\alpha=25\degree $ case evaluated at $x^\star=2$ (left), $x^\star=4$ (middle) and at $x^\star=6$ (right). Solid blue line: contour of the porous disc, dashed blue line: cylindrical mast.}
                    \label{fig:swirl}
                    \end{figure}

\noindent Let us start by assessing the level of swirling motion injected in the wake of the modified porous discs. Fig. \ref{fig:swirl} shows the streamwise evolution ($x^\star = 2$, 4 and 6) of the dimensionless mean swirling velocity $W^\star$ for $\alpha = 15\degree$ and $\alpha = 25\degree$ in comparison to the porous disc with non-pitched blades. For the latter, it is evident that the swirling motion is marginal, aside from a small area where the wake of the disc merges with that of the mast. This is likely due to 3D effects in this wake interaction region. Furthermore, the absence of swirl for the reference case confirms that the porous disc is properly aligned with the direction normal to the incoming flow. As evidenced in Figs. \ref{fig:swirl}(c)-(h), incorporating a pitching angle yields the generation of a swirling motion whose distribution is toroidal. Moreover, the swirl intensity decays as the streamwise distance from the disc increases. Besides, its distribution spreads radially as a consequence of the conservation of the initial angular momentum.

\begin{figure}
            \centering
            \begin{subfigure}{0.46\textwidth}
                \centerline{\includegraphics[width=0.99\textwidth]{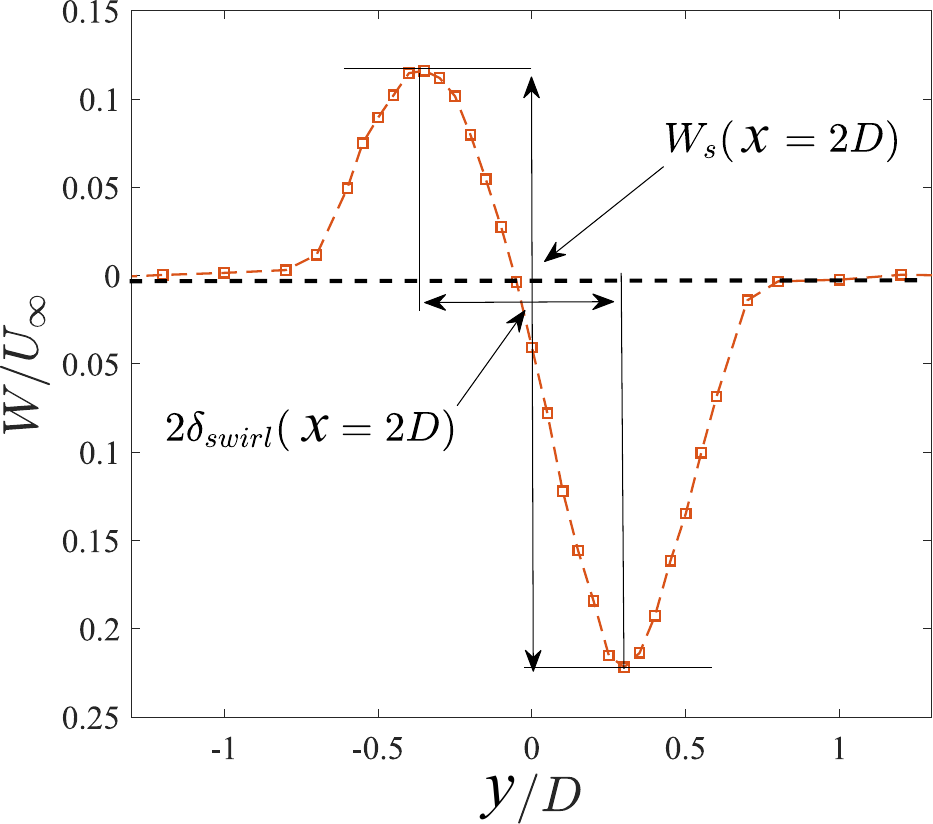}}
                \subcaption{}
                \label{fig:Wmax_pp_defa}
            \end{subfigure}
            \begin{subfigure}{0.46\textwidth}
            \centerline{\includegraphics[width=0.99\textwidth]{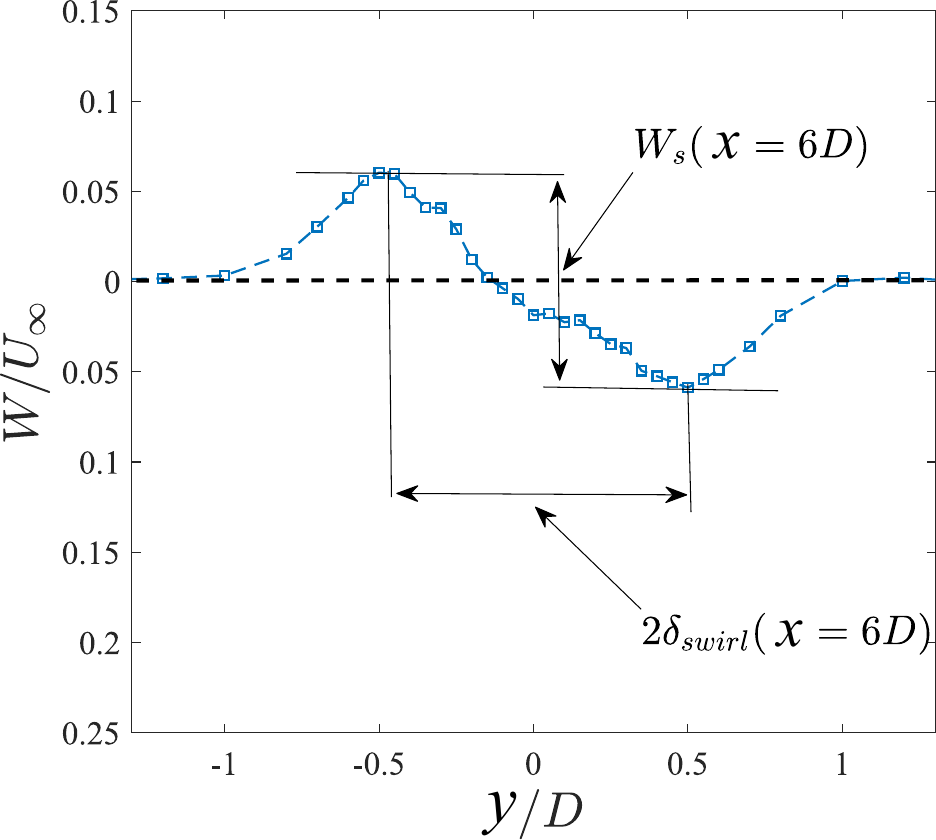}}
            \subcaption{}
            \label{fig:Wmax_pp_defb}
            \end{subfigure}
            \caption{Definition of the peak-to-peak swirling velocity $W_{s}^\star(x)$ and the swirling length $\delta_{swirl}^\star(x)$ for the $\alpha=25\degree $ case at (a) $x^\star=2$ and at (b) $x^\star=6$.}
            \label{fig:Wmax_pp_def}
            \end{figure}

To further characterise the swirling motion, it is essential to derive physical parameters that represent the swirl intensity and its spatial distribution, i.e. $W_{s}$ and $\delta_{swirl}$, respectively. As illustrated in Fig. \ref{fig:Wmax_pp_def}, these quantities are inferred from spanwise profiles of $W^\star$. Note that while the results reported in Fig. \ref{fig:Wmax_pp_def} were obtained at two streamwise locations, $x^\star$ and $x^\star=6$ for $\alpha=25\degree$, profiles with similar shapes have been obtained for the other pitch angles. These profiles confirm that the swirl intensity decays with increasing streamwise distance from the disc. At $x^\star=2$, $W_s^\star$ reaches values of around 0.34, while it falls to around 0.12 at $x^\star=6$. Meanwhile, $\delta_{swirl}^\star$ increases from $0.65$ at $x^\star=2$ to $0.98$ at $x^\star=6$. Furthermore, it is worth noting that the presence of the mast causes an asymmetry on the swirling velocity profiles which is prominent at $X=2D$ but is damped as the wake evolves downstream.

            \begin{figure}
             \centering
             \begin{subfigure}{0.48\textwidth}
                 \includegraphics[width=\textwidth]{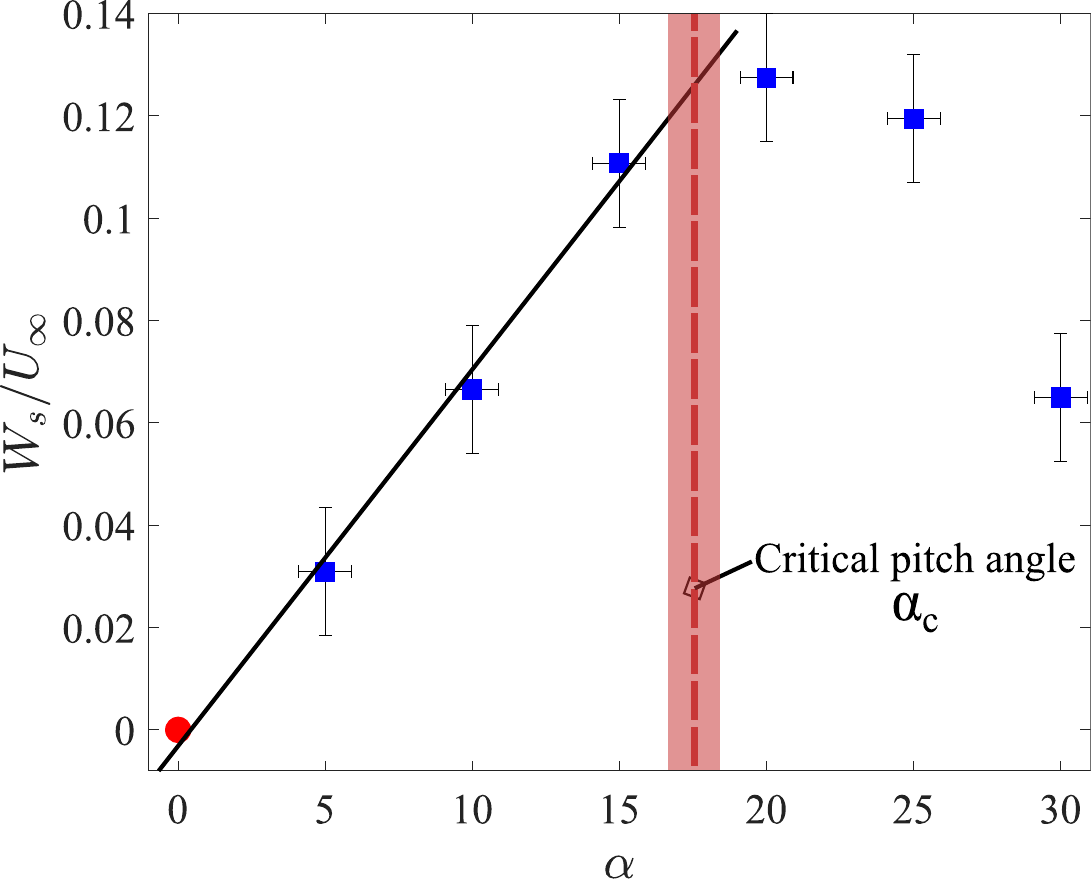}
                 \caption{}
                 \label{fig:W_max_delta_f_alphaa}
             \end{subfigure}
             \begin{subfigure}{0.47\textwidth}
                 \includegraphics[width=\textwidth]{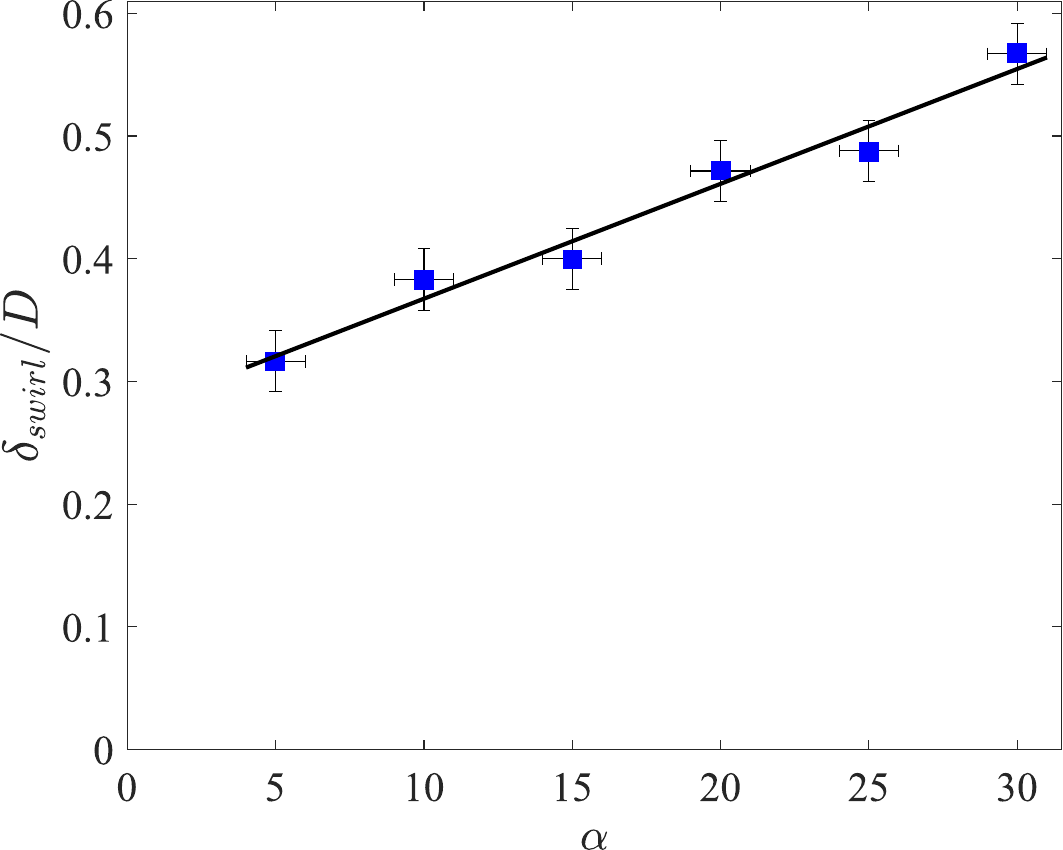}
                 \caption{}
                 \label{fig:W_max_delta_f_alphab}
             \end{subfigure}
              
              \caption{(a) Peak-to-peak swirling velocity $W_s^\star$ and (b) swirling length $\delta_{swirl}^\star$ as a function of the pitch angle $\alpha$ evaluated at $x^\star=6$.}
            \label{fig:W_max_delta_f_alpha}
            \end{figure}

\noindent The effect of the pitch angle on $W_{s}^\star$ and $\delta_{swirl}^\star$ at $x^\star=6$ is reported in Figs. \ref{fig:W_max_delta_f_alphaa} and \ref{fig:W_max_delta_f_alphab}, respectively. The swirl magnitude $W_s^\star$ increases linearly with $\alpha$ until $\alpha=20\degree$ and then drops. This result agrees fairly well with the critical angle $\alpha_c$ at which stall is likely to occur (see \S \ref{subsec:passive_swirl_gen}). This phenomenon represents an intrinsic limitation to the generation of swirl using this kind of modified porous disc. Surprisingly, as emphasised in Fig. \ref{fig:W_max_delta_f_alphab}, the influence of stall is not observed on the evolution of $\delta_{swirl}^\star$ with respect to $\alpha$. The trend shows that the data is well approximated by a linear fit such as $\delta_{swirl}^\star = 0.54 (\alpha - \alpha_0)$, with $\alpha$ in radians and $\alpha_0 = -0.5$ rad. Interestingly, extrapolated to $\alpha=0\degree$, the swirling length is non-null, corresponding to the position of the inner rim of the porous disc. This might provide a potential control parameter to tune the initial swirl length scale. This issue is beyond the scope of this study and is therefore left for future work. In the following, two key metrics of the similarity analysis are investigated: the drag coefficient and the swirl number. 

\subsection{Aerodynamic performances}

                \begin{figure}
                  \centering
                  \includegraphics[width=0.8\textwidth]{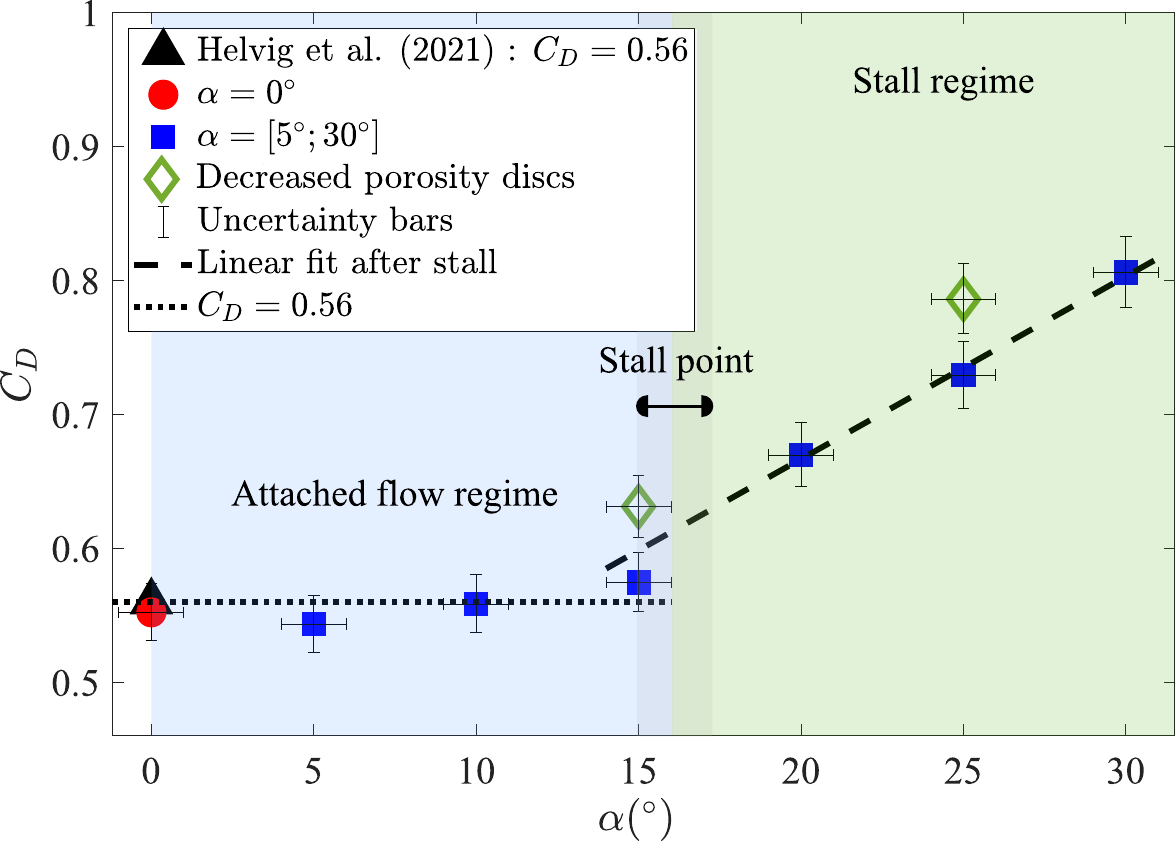}
                  \caption{Measured drag coefficient $C_D$ as a function of $\alpha$. Dotted black line: constant $C_D$ pre-stall, dashed black line: linear trend post-stall.}
                \label{fig:drag_coeff}
                \end{figure}

\noindent The evolution of the drag coefficient $C_D$ with respect to the pitch angle $\alpha$ is displayed in figure \ref{fig:drag_coeff}. For comparison purposes, the value obtained by \cite{helvig_2021}, is also reported. One can remark a very good agreement between our result and that reported in \cite{helvig_2021} although the Reynolds number in this study is much larger. The drag coefficient appears to follow two different trends, with a transition around $\alpha=15\degree$, i.e. a pitch angle slightly lower than the critical value observed for the swirl intensity (see Fig. \ref{fig:W_max_delta_f_alphaa})(a). Indeed, for pitch angles $\alpha \leq 15\degree$, the drag coefficient remains roughly constant, while for $\alpha > 15\degree $, $C_D$ increases linearly with $\alpha$ such as $C_D =0.78\alpha + 0.40$ (with $\alpha$ in radians). As discussed in \S \ref{subsec:passive_swirl_gen}, the onset of stall is expected to cause a reduction in the effective porosity of the discs. To confirm this assumption, the results obtained for low-porosity discs without swirl (see \S \ref{subsec:passive_swirl_gen} for details) are also plotted in Fig. \ref{fig:drag_coeff}. Comparing these low-porosity discs to their counterparts at $\alpha = 15\degree$ and $25\degree$, one can clearly see that reducing the effective porosity without injecting azimuthal momentum leads to an increase in drag comparable to that caused in the post-stall regime for discs with pitched blades. While in both cases the increase in drag is well adjusted by a linear fit with the same slope, it can be noticed that the drag of the low-porosity discs is approximately 8\% higher relative to their stalled counterparts. This positive offset is probably due to a contraction effect of the streamlines as the working fluid passes around the blades, which tends to reduce even more the effective porosity.
           
\noindent Although the tip speed ratio (TSR) is used to characterise the rotational motion in wind turbine applications, this parameter is of little relevance in this study, as the vortex is generated in a stationary manner. Instead, the swirl number $\Hat{S}$ will be used to effectively quantify and compare the swirl generated in the wake of the modified porous discs to that of wind turbines. Following \cite{reynolds1962similarity} and \cite{alekseenko1999helical}, the swirl number $\Hat{S}$ can be calculated as follows 

            \begin{equation}
                \Hat{S} = \frac{\int_0^\infty ( U^\star W^\star + \overline{u'w'}^\star) r^{\star^2}dr^\star}{\int_0^\infty \left[ U^\star \Delta U^\star + \frac{W^{\star^2}}{2} - \overline{u'^2}^\star + \frac{\overline{w'^2}^\star +\overline{v'^2}^\star}{2}\right] r^\star dr^\star}. 
                \label{eq:swirl_number_gupta}
            \end{equation}
            
\noindent To the authors knowledge, very few studies have reported on the swirl number of real-scale wind turbines, and even fewer on establishing a correlation between the TSR and the swirl number. For this reason, only estimates of $\Hat{S}$ for wind turbines exist in current literature \citep{lee2020experimental,Wosnik2013,morris2016evaluation,holmes2022impact,bortolotti2019iea}, giving values of swirl $\Hat{S} \in [0 , 0.3]$. Furthermore, it is worth noting that our measurements emphasise that the predominant terms in Eq. (\ref{eq:swirl_number_gupta}) are the mean shear and the momentum deficit flowrate meaning that the swirl number is well approximated by

\begin{equation}
                \Hat{S} \approx \frac{\int_0^\infty U^\star W^\star r^{\star^2}dr^\star}{\int_0^\infty U^\star \Delta U^\star r^\star dr^\star}. 
                \label{eq:swirl_number_gupta_approx}
            \end{equation}
            
            \begin{figure}
                \centering
                \includegraphics[width=0.75\textwidth]{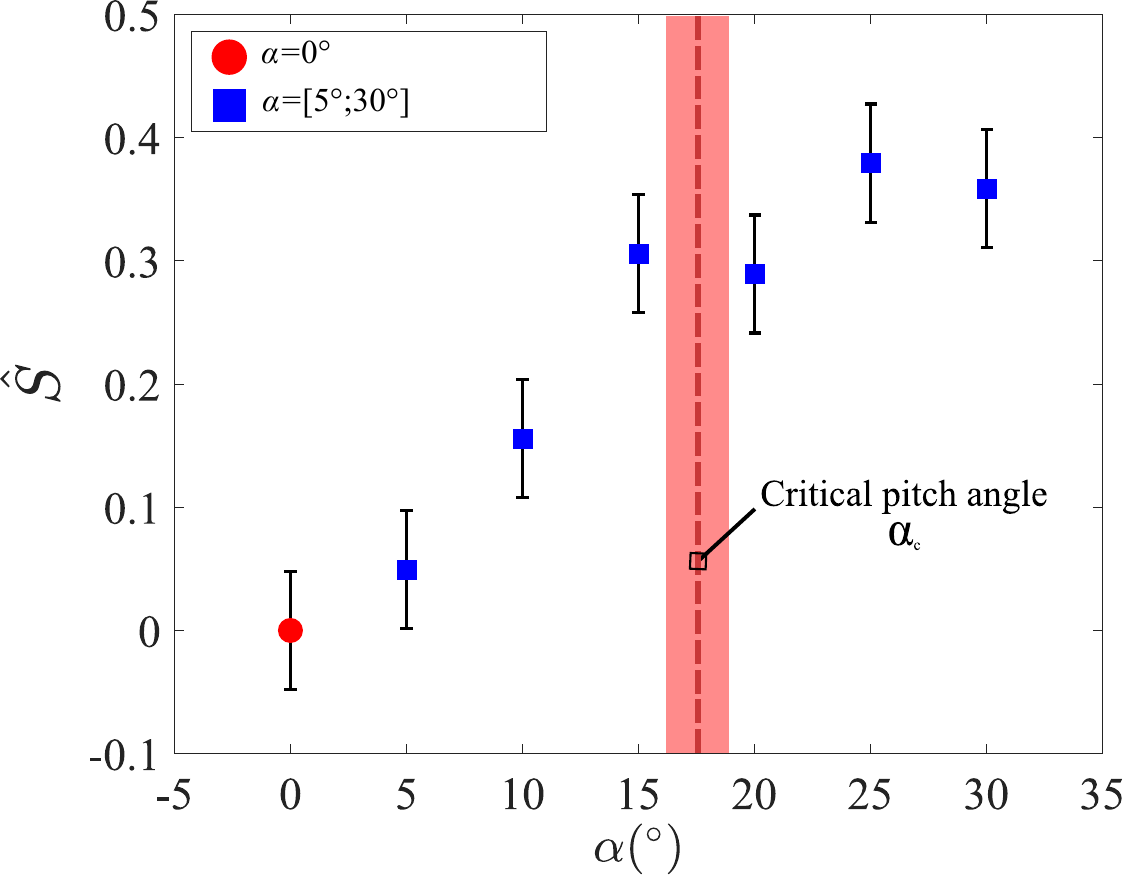}
                \caption{Swirl number $\Hat{S}$ as a function of the pitch angle $\alpha$ measured at $x^\star=6$.}
                \label{fig:S_f_alpha}
            \end{figure}

\noindent The effect of the pitch angle on the estimated swirl number $\Hat{S}$ is emphasised in figure \ref{fig:S_f_alpha}. Before the critical pitch angle $\alpha_c$, the swirl number increases with $\alpha$ until reaching a plateau in the post-stall regime. Remarkably, prior to the onset of stall, the swirl number generated by the modified porous discs falls within the range of values observed in real-scale wind turbines, meaning that the new design proposed in this study enables to reproduce swirl intensity comparable to that encountered at all stages of wind turbine operation without altering the value of $C_D$ \citep{bortolotti2019iea,holmes2022impact}.

\noindent In conclusion to this section, it appears that pitching the blades of a porous disc is an effective and inexpensive way of reproducing the aerodynamic properties, at least on a macroscopic scale, of a wind turbine. Two characteristic angles around the critical stall angle emerge from this first analysis. More specifically, these angles are $\alpha = 15\degree$ and $\alpha = 25\degree$ and will be used in the following to benchmark the influence of swirl and porosity respectively. The test case $\alpha = 15\degree$ corresponds to a regime where the blades have not yet stalled and for which the swirl intensity is almost at its maximum. Moreover, at this angle, the drag coefficient is equal to that of the reference case (i.e. $\alpha = 0\degree$). This implies that data collected at $\alpha = 15\degree$ can therefore be used to isolate the effect of the swirling motion on the evolution of the wake. On the other hand, the test case $\alpha = 25\degree$ corresponds to a flow in the post-stall regime, but for which the generated swirl intensity is comparable to that obtained at $\alpha = 15\degree$. However, the drag coefficient featuring $\alpha = 25\degree$ is greatly increased due to the reduction in apparent porosity. The comparison between the cases $\alpha = 15\degree$ and $\alpha = 25\degree$ will therefore make it possible to isolate the influence of effective porosity on the evolution of the swirling wake.

\section{Mean wake survey}

\noindent In this section, an in-depth analysis of the wake evolution is provided. As previously explained, the investigation is conducted for cases $\alpha=15\degree $ and $\alpha=25\degree $ alongside with the reference case $\alpha=0\degree$ to emphasise the role of swirl.

\subsection{Mean flow topology}

            \begin{figure}
              \centering
              \begin{subfigure}{0.98\textwidth}
              \centerline{\includegraphics[width=\textwidth]{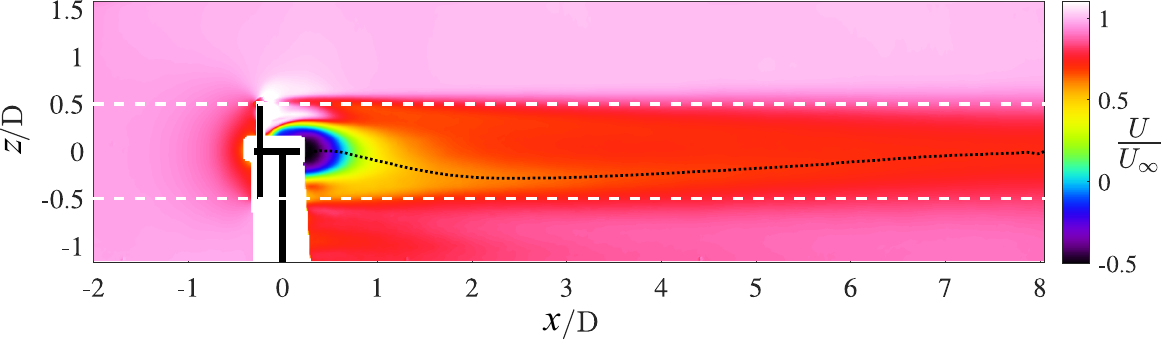}}
              \subcaption{}
              \label{fig:vel_def_PIVa}
              \end{subfigure}
              
              \begin{subfigure}{0.98\textwidth}
              \centerline{\includegraphics[width=\textwidth]{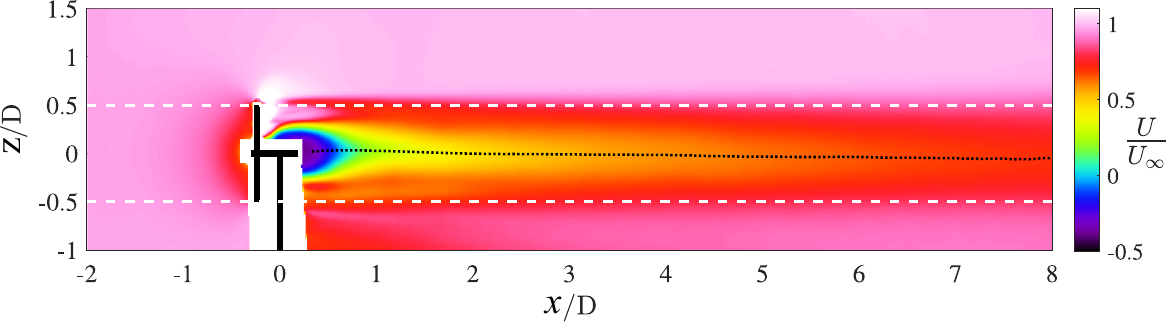}}
                 \subcaption{}
                 \label{fig:vel_def_PIVb}
              \end{subfigure}
   
               \begin{subfigure}{0.98\textwidth}
              \centerline{\includegraphics[width=\textwidth]{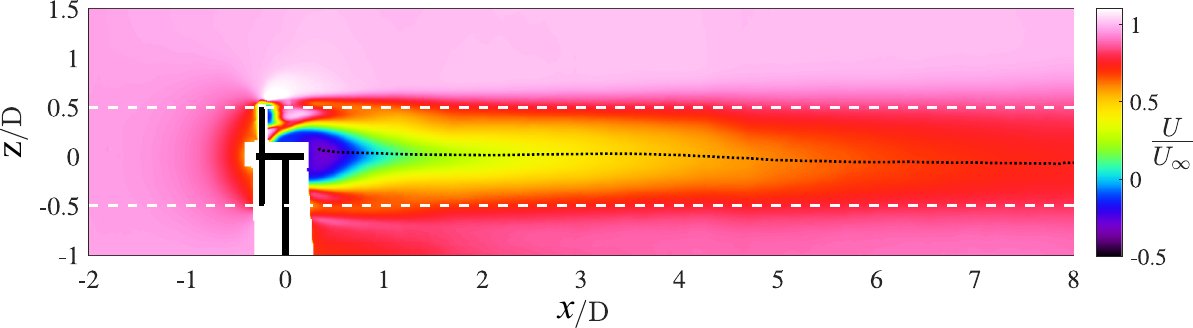}}
                \subcaption{}
                \label{fig:vel_def_PIVc}
              \end{subfigure}

              \caption{Normalised streamwise velocity $U^\star$ fields for the (a) $\alpha=0\degree $ case, (b) the $\alpha=15\degree $ case and for (c) the $\alpha=25\degree $ case. The white area near $x^\star = 0$ is the masked zone where the test rig (black patched area) is located. Dashed white lines: edges of the porous disc, black dotted lines: wake centre coordinates.}
            \label{fig:vel_def_PIV}
            \end{figure}

\begin{figure}
	\centering
	\includegraphics[width=0.8\textwidth]{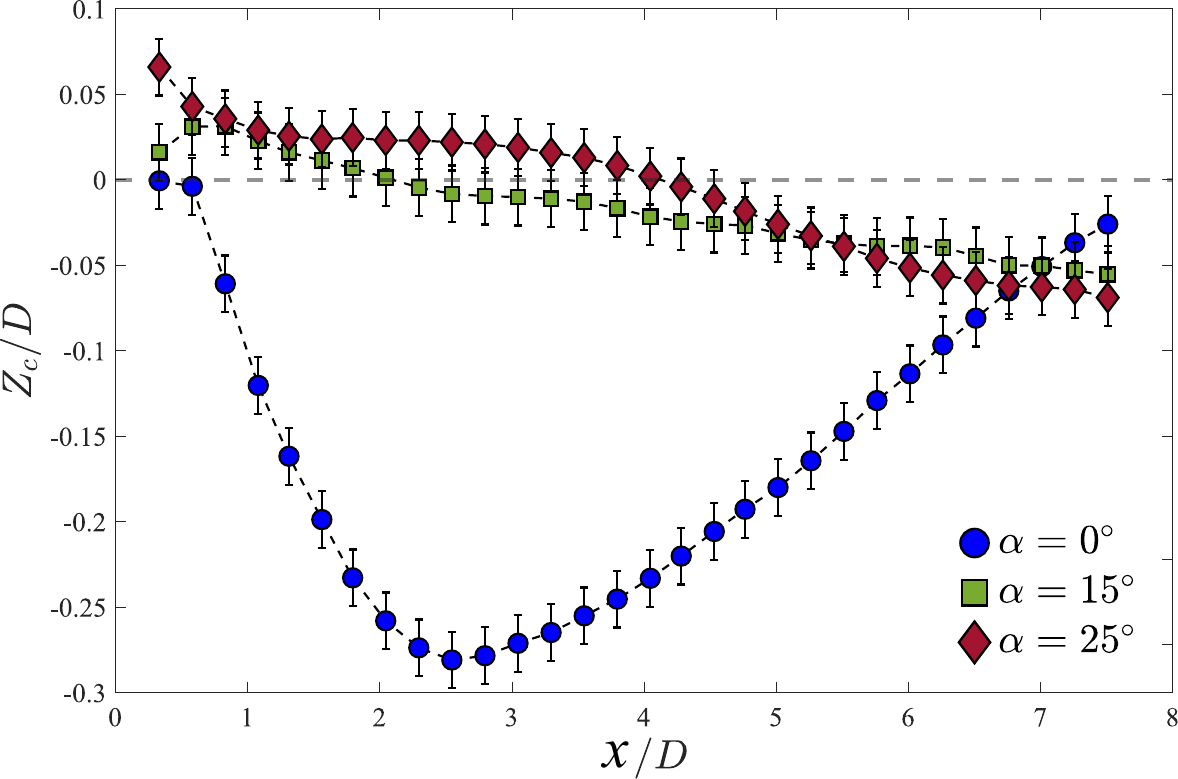}
	\caption{Streamwise evolution of the mean wake centre vertical coordinate $Z_c^\star(x)$ for $\alpha=[0\degree, 15\degree, 25\degree]$. Only 1/15 of the markers are plotted for visibility.}
	\label{fig:wake_centre_y0}
\end{figure}

\begin{figure}
                     \centering
                     \includegraphics[width=0.9\textwidth]{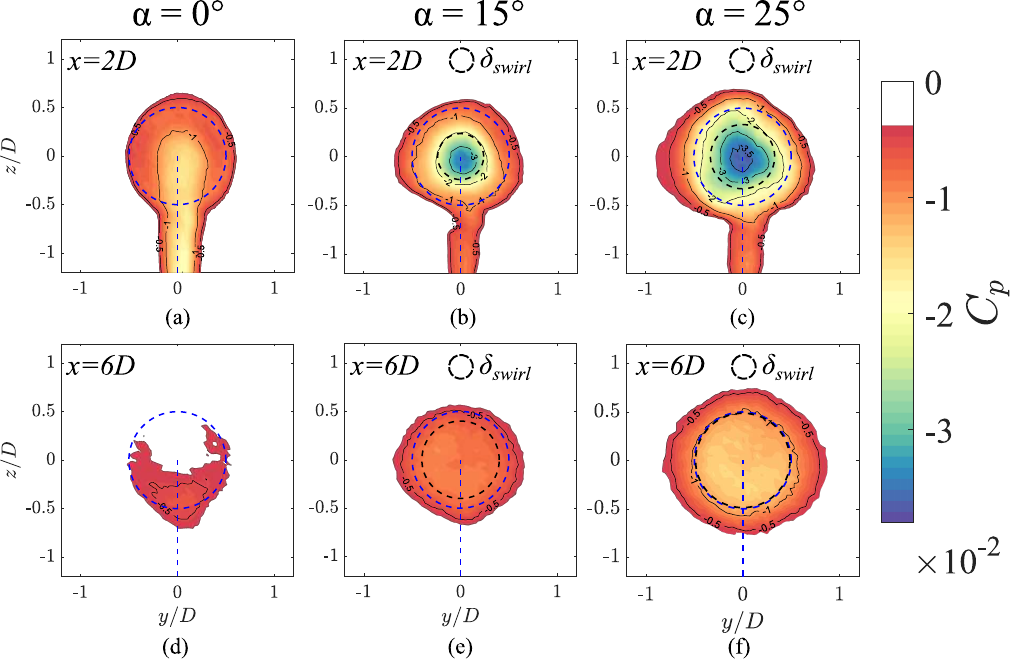}
  
                  \caption{Reconstructed pressure coefficient $C_P$ spanwise planes for the $\alpha=0\degree $ case (left), the $\alpha=15\degree $ case (middle) and for the $\alpha=25\degree $ case (right) evaluated at $x^\star=2$ (a,b,c) and at $x^\star=6$ (d,e,f). Dashed black lines: swirl vortex core, dashed blue lines: test rig outline.}
                \label{fig:CP_HWAMAP}
                \end{figure}

\noindent The mean streamwise velocity fields for cases $\alpha=[0\degree, 15\degree, 25\degree]$ along the streamwise ($y^\star=0$) plane are reported in Fig. \ref{fig:vel_def_PIV}. For each case, the very near wake ($x^\star \leq 1$) is characterised by the presence of a recirculation region, which is likely caused by the central solid disc at hub-height of each surrogate (see Fig. \ref{fig:PD_GP}). The presence of the porous disc yields a velocity deficit area which remains roughly confined within the disc region until $x^\star=8$. In addition, the cylindrical mast produces its own wake, which is observable up to approximately $x^\star=2$, meaning that strong interactions between the disc and the mast are expected at the early stage of the wake development. To assess the influence of this wake interaction on the mean flow topology, Fig. \ref{fig:wake_centre_y0} displays the streamwise evolution of the dimensionless vertical position $Z_c^\star$ of the wake centre, which is assimilated to the location the minimum streamwise velocity. Regarding the reference disc ($\alpha=0\degree$), it appears clearly that the wake centre is deviated downwards until $x^\star=2.5$, where it reaches a minimum height $Z_c/D=-0.3$, and then recovers towards hub-height further downwind ($x^\star=7$). This downwash motion is a direct consequence of the pressure drop caused by the mast's wake as emphasised in Fig. \ref{fig:CP_HWAMAP}(a), which displays the mean pressure coefficient $C_p$ ($\equiv 2 (P - P_\infty)/\rho U_\infty^2$ where $P$ is the mean pressure, $P_\infty$ is the free-stream pressure and $\rho$ the air density) which is inferred from the velocity measurements following the methodology developed in \cite{shanmughan2020optimal}. Indeed, the presence of the mast induces a low pressure region which pulls the wake downwards. Further downstream, since the wake recovers, the imprint of the mast on the pressure field diminishes, allowing the wake to stabilise at hub-height.

\noindent Given that the radial pressure gradient is primarily balanced by the swirling motion \citep{Shiri2010}, the introduction of swirl is expected to influence the pressure distribution and, consequently, alter the mean flow topology. This is well supported by the results reported in Fig. \ref{fig:wake_centre_y0}, which show that the wake generated by the pitched porous discs is only slightly deflected downwards, regardless of the value of $\alpha$. The impact of swirl on the pressure distribution is emphasised in Figs. \ref{fig:CP_HWAMAP}(b) and \ref{fig:CP_HWAMAP}(c) for $\alpha = 15\degree$ and $\alpha = 25\degree$, respectively. In both cases, while the imprint of the mast wake is still visible, the corresponding pressure coefficient is much larger ($C_p \approx -0.75 \times 10^{-2}$) than that of the reference disc ($C_p \approx -1.5 \times 10^{-2}$). More interestingly, unlike the non-pitched case, the pressure distribution downstream of the pitched discs is nearly axisymmetric, with a pronounced low pressure core centred at hub height. This low pressure core counterbalances the suction induced by the mast wake, enabling the disc wake to resist downwash.

                \begin{figure}
                   \centering
        
                   \includegraphics[width=0.9\textwidth]{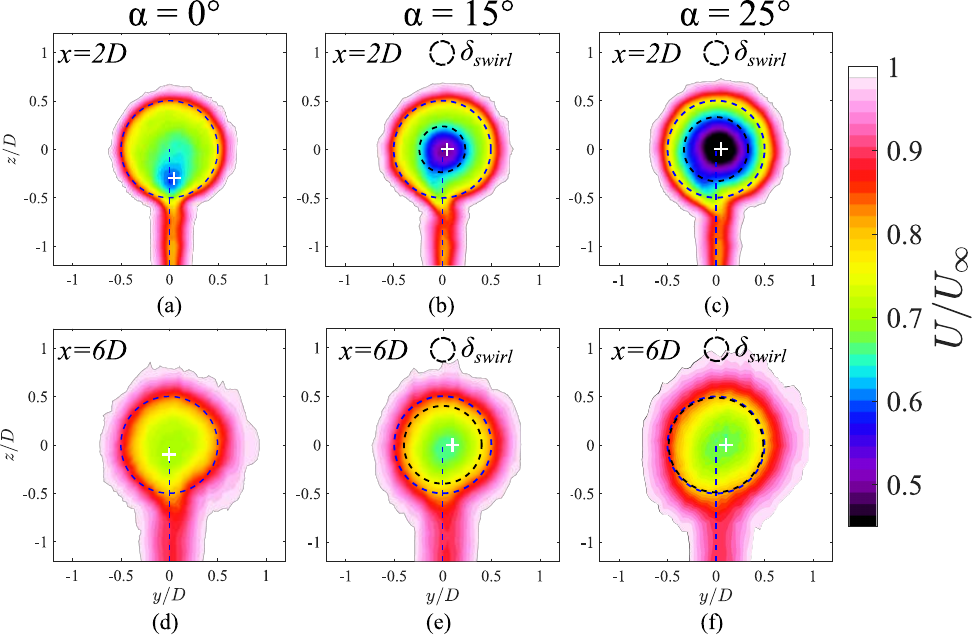}
  
                  \caption{Normalised streamwise velocity $U^\star$ spanwise planes for the $\alpha=0\degree $ case (left), the $\alpha=15\degree $ case (middle) and for the $\alpha=25\degree $ case (right) evaluated at $x^\star=2$ (a,b,c) and at $x^\star=6$ (d,e,f). Solid blue line: porous disc edges, dashed blue line: mast outline, white plus marker: wake centre.}
                \label{fig:Umean_HWAMAP}
                \end{figure}    

\noindent Axisymmetry being a key ingredient in the self-similarity framework, we now investigate how the change in initial conditions affects the spanwise distribution of mean streamwise velocity.  In the vicinity of the disc, the wake deflection induced by the mast for the reference case (i.e. $\alpha = 0\degree$) is clearly evident in Fig. \ref{fig:Umean_HWAMAP}(a), causing an axisymmetry breaking. On the contrary, Fig. \ref{fig:Umean_HWAMAP}(b) emphasises that adding swirl without altering the effective porosity generates a nearly axisymmetric wake deficit. Moreover, the velocity deficit obtained for $\alpha = 15\degree$ is much higher than that featuring the reference disc at the same streamwise position. As a complementary remark, one can see that the radius of the central core of velocity defect is comparable to that of the low-pressure core (see Fig. \ref{fig:CP_HWAMAP}(b)). Overall, these results suggest that, through the action of pressure, the swirling motion significantly affects the velocity distribution at the early stage of wake development. As shown in Fig. \ref{fig:Umean_HWAMAP}(c), increasing the effective porosity yields similar conclusions, albeit an increase in the maximum velocity deficit, reflecting the drag increase observed in the post-stall regime. Figs. \ref{fig:Umean_HWAMAP}(d), \ref{fig:Umean_HWAMAP}(e) and \ref{fig:Umean_HWAMAP}(f) illustrate that, further downstream, the differences between the non-swirling wake and the swirling wake vanish, probably due to the decay in swirl intensity. 

\noindent All in all, these findings provide evidence that while the change in apparent porosity appears to a have marginal influence on flow topology, the addition of swirl significantly impacts the initial conditions on which the wake of the actuator disc develops. Furthermore, a close inspection of the low-pressure core shown in Figs. \ref{fig:CP_HWAMAP}(b) and \ref{fig:CP_HWAMAP}(c) reveals that its radial extension is set by $\delta_{swirl}$, indicating that this length scale plays a major role in the early stage of the wake development.

\subsection{Scaling laws of the swirling wake's properties}

 \begin{figure}
             \centering
             \begin{subfigure}{0.48\textwidth}
                 \includegraphics[width=\textwidth]{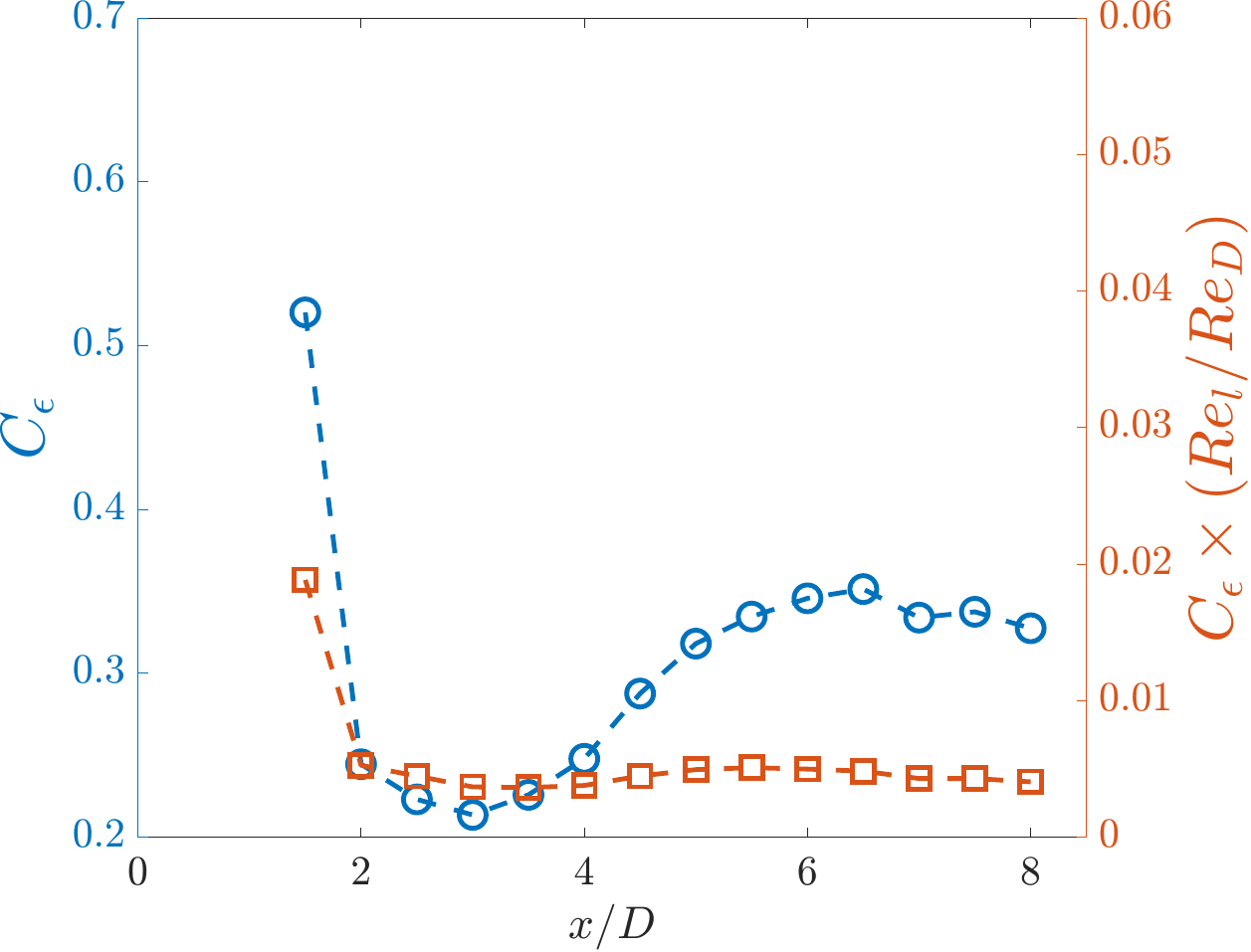}
                 \caption{}
                 \label{fig:C_epsilon_f_x_s0}
             \end{subfigure}
             \begin{subfigure}{0.48\textwidth}
                 \includegraphics[width=\textwidth]{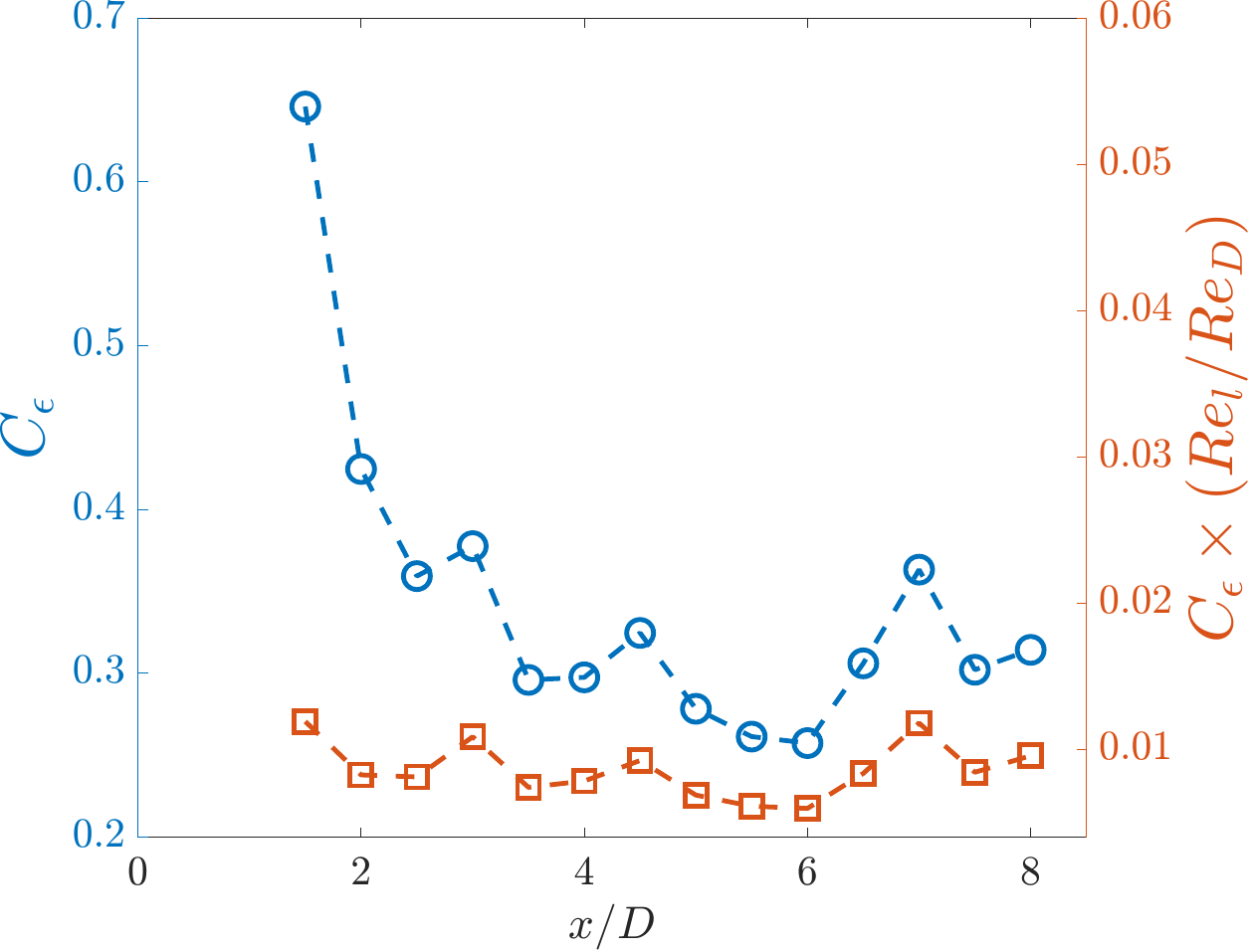}
                 \caption{}
                 \label{fig:C_epsilon_f_x_s15}
             \end{subfigure}
              
              \caption{Streamwise evolution of the dissipation coefficient $C_\epsilon$ and of the product $(Re_l/Re_D)\times C_\epsilon$ calculated at the centreline ($y^\star=z^\star=0$) for cases (a) $\alpha=0\degree$ and (b) $\alpha=15\degree$.}
            \label{fig:C_epsilon_f_x}
            \end{figure}

\noindent Echoing the discussions made in \S \ref{sec:similarity}, we first examine whether swirl changes the nature of turbulence in the wake (equilibrium vs. non-equilibrium). To this end, the centreline streamwise evolution of the dissipation coefficient $C_\epsilon(x)$ and of the product $(Re_l(x)/Re_D)\times C_\epsilon(x)$ is plotted in Fig. \ref{fig:C_epsilon_f_x} for cases $\alpha=0\degree$ and $\alpha=15\degree$. In both scenarios, the pre-multiplied dissipation coefficient $(Re_l(x)/Re_D)\times C_\epsilon(x)$ exhibits a plateau beyond $x^\star=2$ whereas $C_\epsilon(x)$ does not. Similar trends are observed for the $\alpha=25\degree$ case (not shown here). This result shows that non-equilibrium turbulence is a more relevant assumption for both the swirling and non-swirling wakes. It is therefore argued that the swirling motion does not change the nature of the turbulent flow but sets the length and velocity scales that drive self-similarity. To complete this assessment, a scaling analysis of the mean wake properties is carried out in the following section. In particular, we check if the novel mean swirl decay scaling law Eq. ($\ref{eq:Ws_novel_scaling}$), derived in $\S\ref{sec:similarity}$, is present in our swirling wake. The methodology used to determine the best fit parameters for each law is similar to the non-linear fit methods used by \cite{nedic2013fractal} and later improved in the work of \cite{dairay_non-equilibrium_2015}. The following functions are considered for the scaling laws:

                \begin{align}
                    \frac{U_s}{U_\infty} = & \ A\left( \frac{x-x_{0_{U}}}{\theta}\right)^{\beta_{U}}, \label{eq:Us_fit_scaling}\\[12pt]
                    \frac{\delta_{swirl}}{\theta} = & \ B\left( \frac{x-x_{0_{\delta}}}{\theta}\right)^{\beta_{\delta}},  \label{eq:delta_fit_scaling}\\[12pt]
                    \frac{W_s}{U_\infty} = & \ C\left( \frac{x-x_{0_{W}}}{\theta}\right)^{\beta_{W}}, \label{eq:Ws_fit_scaling} 
                \end{align}

\noindent where $\beta_{U}$, $\beta_{\delta}$ and $\beta_{W}$ are the fit exponents of the power laws for the velocity deficit, the swirling length and the swirling velocity, respectively. $x_{0_U}$ , $x_{0_\delta}$ and $x_{0_W}$ are the corresponding virtual origins, A, B and C are fit constants and $\theta$ is the momentum thickness. The first step of the fitting method consists in setting all virtual origins to 0 and performing initial linear fits of $(U_s/U_\infty)^{1/\beta_U}$, $(\delta_{swirl}/\theta)^{1/\beta_{\delta}}$ and $(W_s/U_\infty)^{1/\beta_{W}}$ in order to obtain a first approximation for the power law exponents. These values are then used in order to bound and initialise the nonlinear least-squares regression algorithm. This algorithm is then used to simultaneously find the optimal values for the parameters. The resulting parameters from the fitting method are listed in table \ref{tab:fits}.

                \begin{table}
                  \begin{center}
                \def~{\hphantom{0}}
                  \begin{tabular}{lcccc}
                      Case  & function & $A,B,C$   &   $\beta_{(U,\delta,W)}$ & $x_{0_{(U,\delta, W)}}$ \\[3pt]
                       $\alpha=15\degree $ & $U_s/U_\infty$ & 11.99 & ~~-0.9~ & -23.6\\
                       $\alpha=15\degree $ & $\delta_{swirl}/\theta$ & 0.65 & ~~0.54 & -0.50\\
                       $\alpha=15\degree $ & $W_s/U_\infty$ & 19.9 & ~~-1.36 & -11.1\\
                       \\
                       \hline \\
                       $\alpha=25\degree $ & $U_s/U_\infty$ & 13.06 & ~-1.02 & -5.7\\
                       $\alpha=25\degree $ & $\delta_{swirl}/\theta$ & 0.96 & 0.47 & -1.38\\
                       $\alpha=25\degree $ & $W_s/U_\infty$ & 21.02 & -1.40 & -8.54\\
                  \end{tabular}
                  \caption{Non-linear fit parameters obtained for the scaling laws (\ref{eq:Us_fit_scaling})-(\ref{eq:Ws_fit_scaling}).}
                  \label{tab:fits}
                  \end{center}
                \end{table}

\begin{figure}
                  \centering
                  \begin{subfigure}{0.48\textwidth}
                  \includegraphics[width=\textwidth]{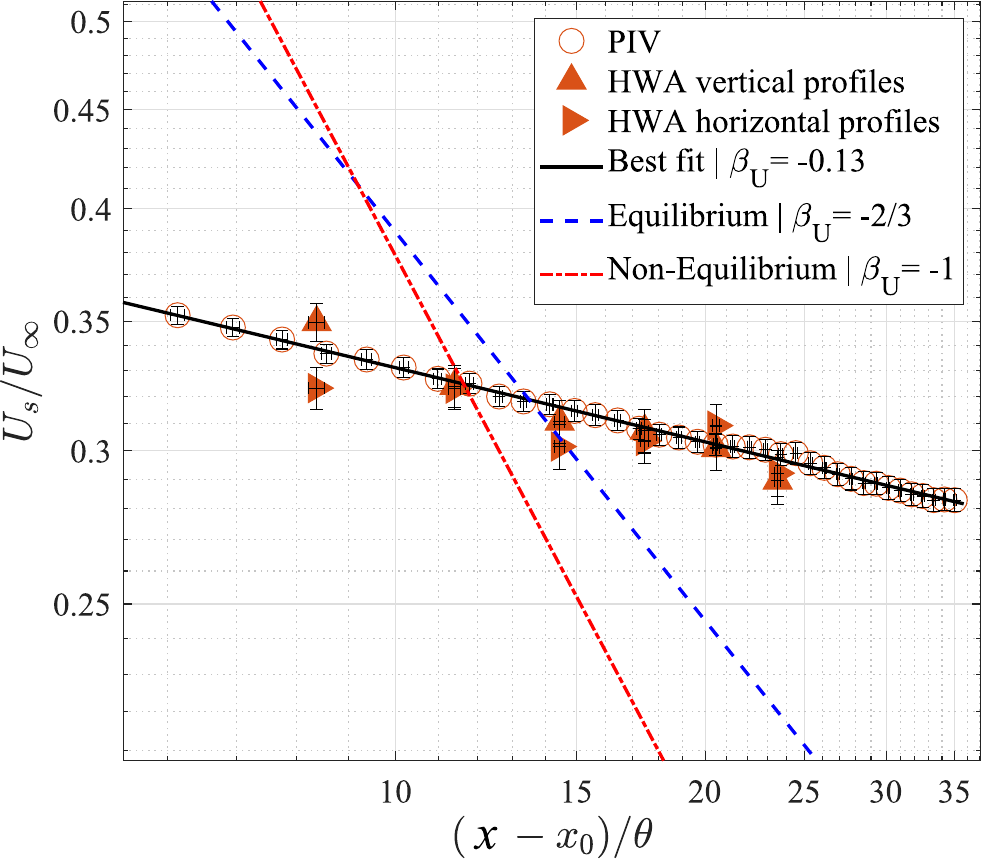}
                  \subcaption{}
                  \label{fig:Us_scalinga}
                  \end{subfigure}

                  \centering
                  \begin{subfigure}{0.48\textwidth}
                  \includegraphics[width=\textwidth]{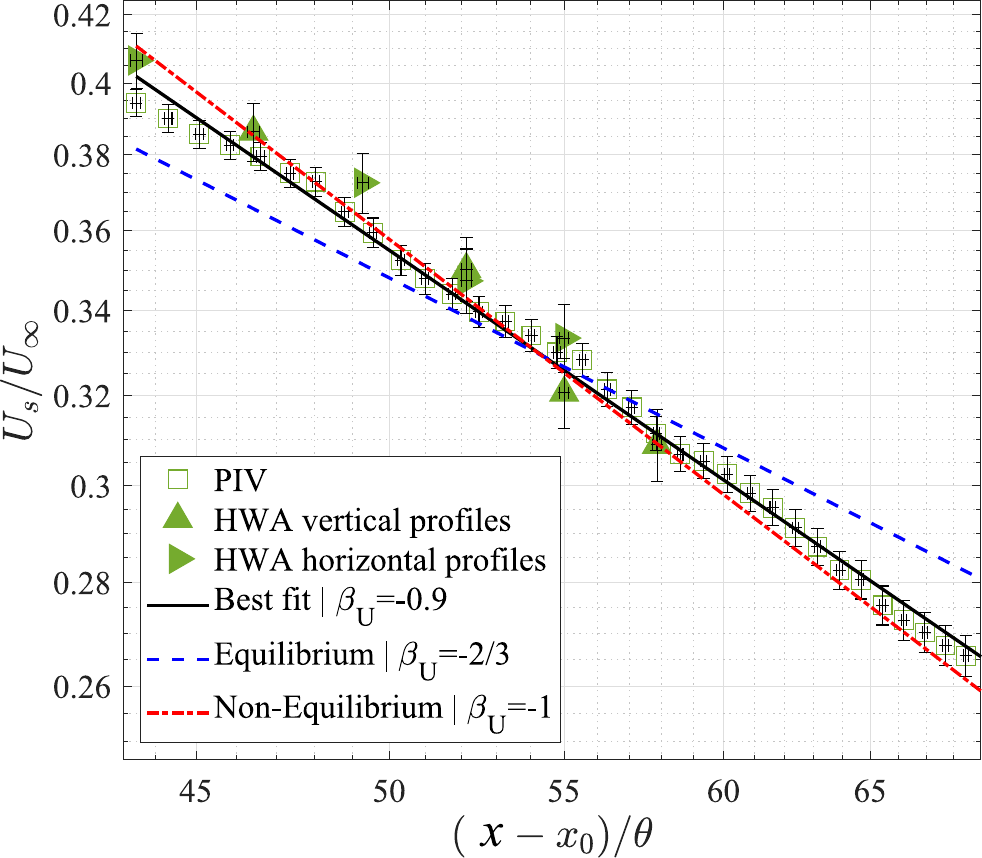}
                  \subcaption{}
                  \label{fig:Us_scalingb}
                  \end{subfigure}
                    \vspace{0.25cm}
                    
                    \centering
                    \begin{subfigure}{0.48\textwidth}
                  \includegraphics[width=\textwidth]{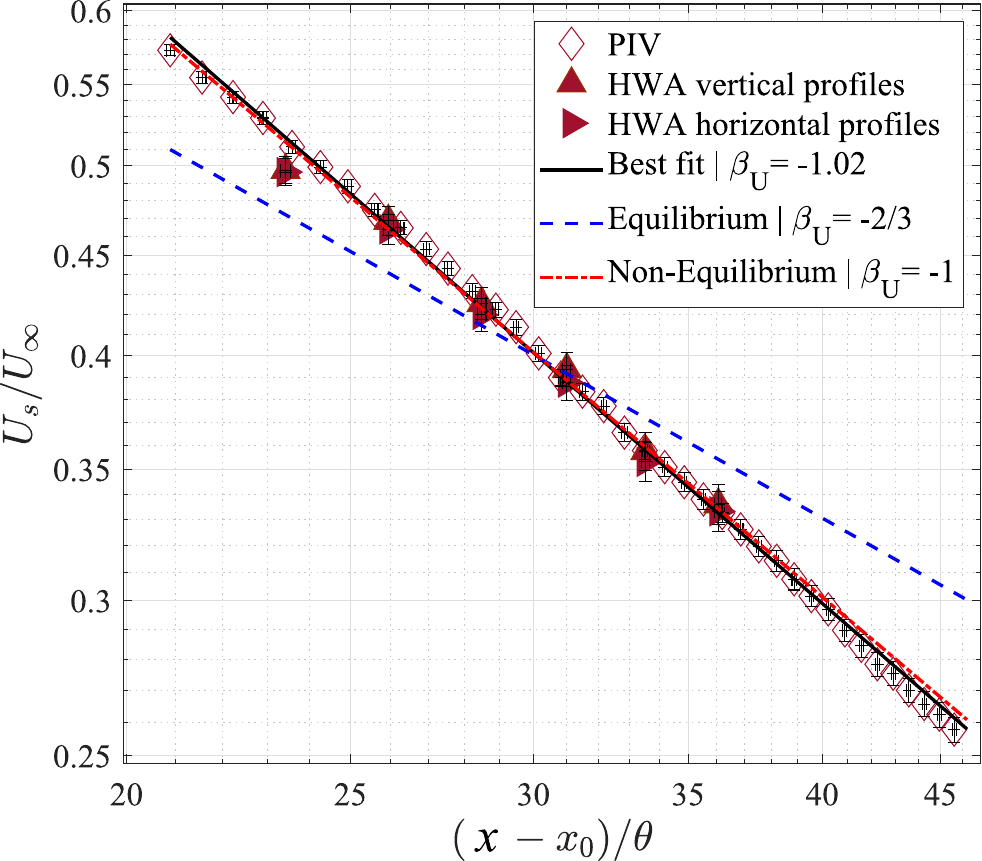}
                  \subcaption{}
                  \label{fig:Us_scalingc}
                  \end{subfigure}

                  \caption{Scaling plots of the characteristic axial velocity deficit $(U_s/U_\infty)$ for (a) the $\alpha=0\degree $ case, for (b) the $\alpha=15\degree $ case and (c) the $\alpha=25\degree $ case. Only 1/8 of the markers are plotted for clarity.}
                \label{fig:Us_scaling}
                \end{figure}

\noindent The streamwise evolution of $U_s(x)/U_\infty$ with the resulting scaling laws for cases $\alpha=15\degree $ and $\alpha=25\degree $ is reported in figure \ref{fig:Us_scaling}. Additionally, the $\alpha=0\degree $ case is also considered in Fig. \ref{fig:Us_scalinga} for reference. Case $\alpha=0\degree $ clearly shows no signs of either similarity scaling law for the considered streamwise distances, and the fitted law has a light slope ($U_s(x) \sim x^{0.13}$). For the other cases, the deficit recovery (figures \ref{fig:Us_scalingb} and \ref{fig:Us_scalingc}) is well predicted by the non-equilibrium predictions ($\beta_{U} = -1$).
                
                \begin{figure}
             \centering
                  \begin{subfigure}{0.455\textwidth}
                
                  \centerline{\includegraphics[width=\textwidth]{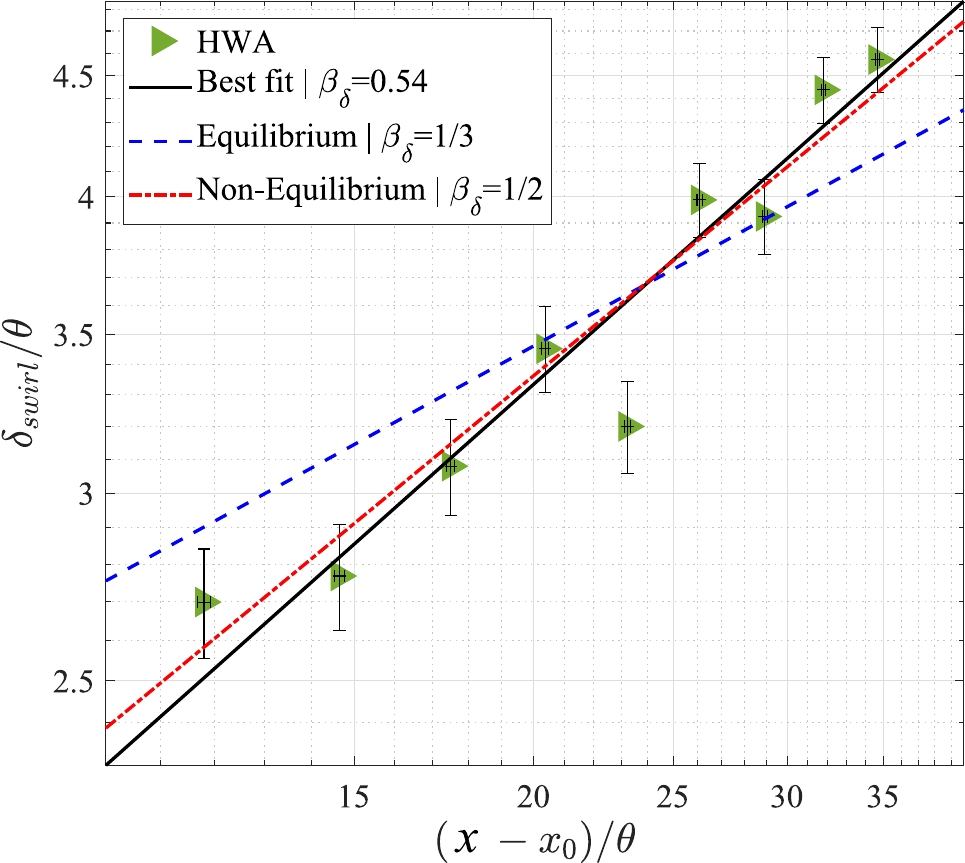}}
                  \subcaption{}
                  \end{subfigure}
                  \begin{subfigure}{0.455\textwidth}
                  \centerline{\includegraphics[width=\textwidth]{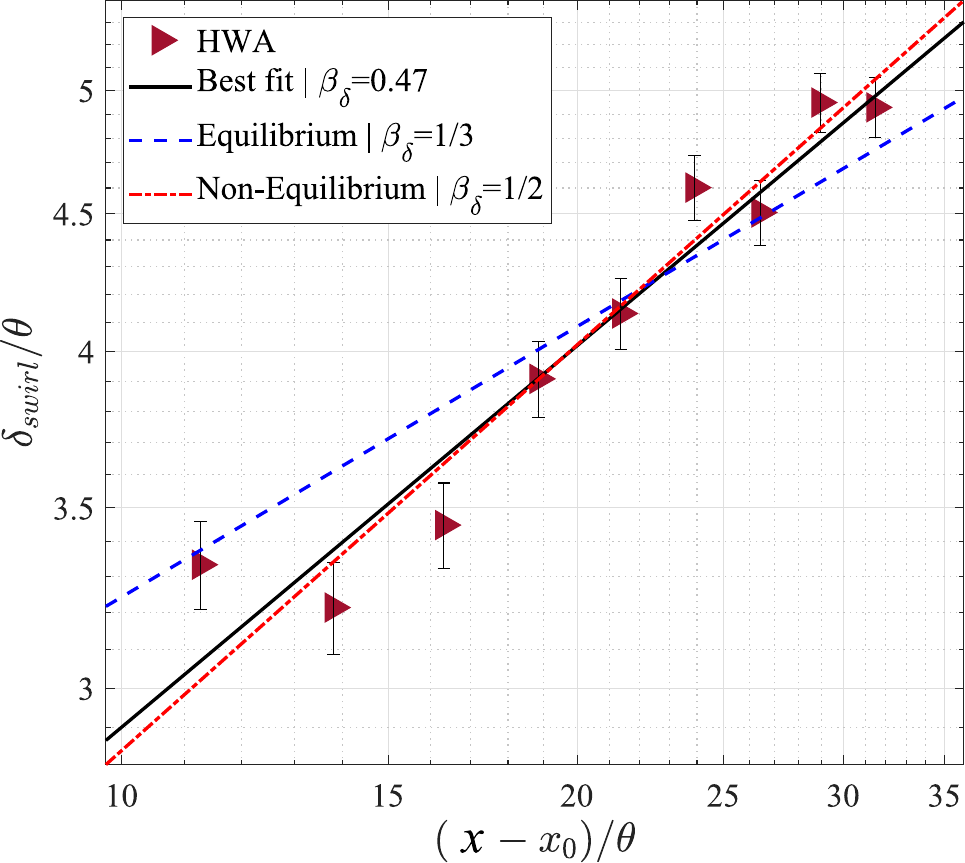}}
                  \subcaption{}
                  \end{subfigure}
  
                  \caption{Scaling plots of the length scale $\delta_{swirl}$ for (a) the $\alpha=15\degree $ case and for (b) the $\alpha=25\degree $ case.}
                \label{fig:delta_swirl_scaling}
            \end{figure}

             \begin{figure}
             \centering
                  \begin{subfigure}{0.46\textwidth}
                  \centerline{\includegraphics[width=\textwidth]{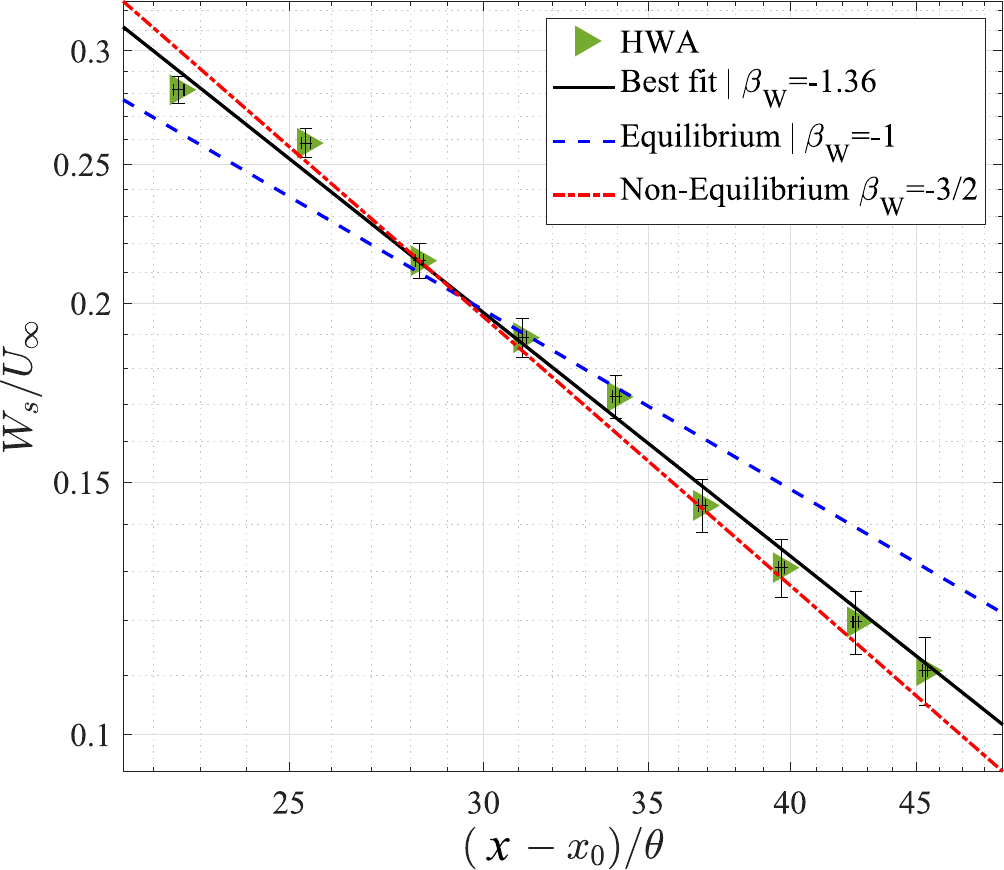}}
                  \subcaption{}
                  \end{subfigure}
                  \begin{subfigure}{0.46\textwidth}
                  \centerline{\includegraphics[width=\textwidth]{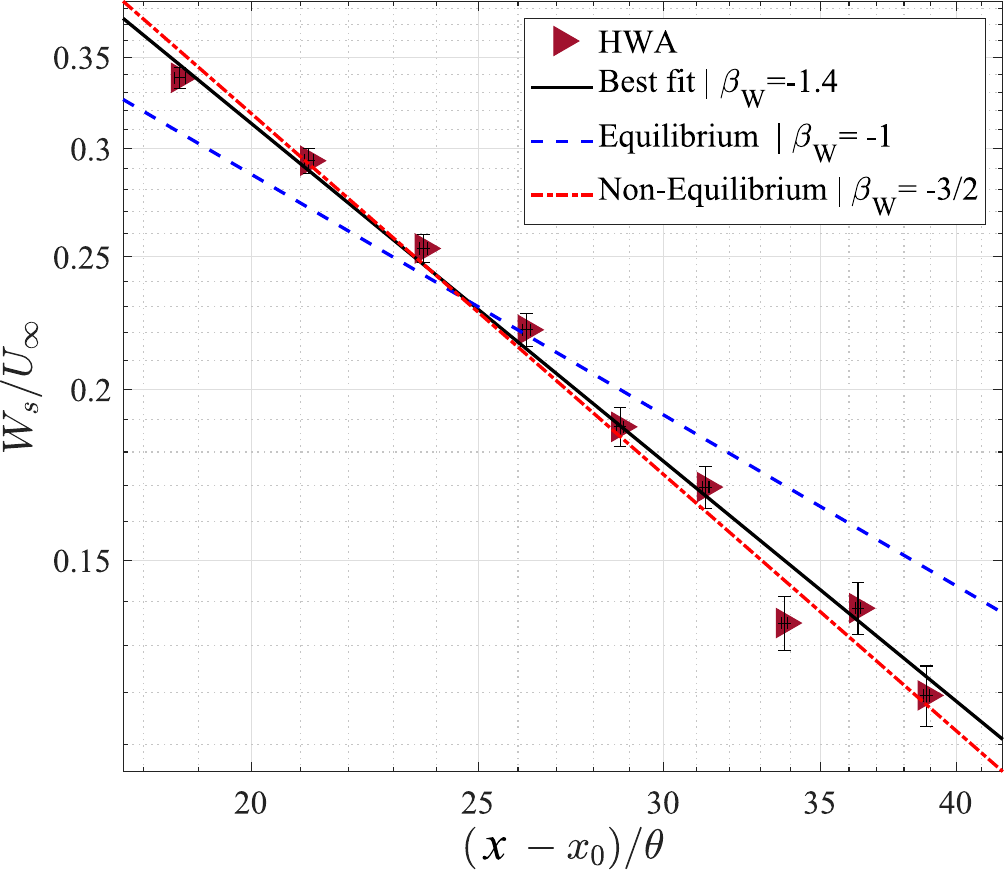}}
                  \subcaption{}
                  \end{subfigure}
                    
                  \caption{Scaling plots of the characteristic swirling velocity for (a) the $\alpha=15\degree $ case and for (b) the $\alpha=25\degree $ case.}
                \label{fig:Ws_scaling}
            \end{figure}

\noindent As for the scaling of $\delta_{swirl}$, Fig. \ref{fig:delta_swirl_scaling} shows the streamwise evolution of the swirling length scale $\delta_{swirl}(x)/\theta$ for both swirling wakes. The results show very good agreement with the non-equilibrium similarity predictions ($\beta_{\delta} = 1/2$) for both cases. Further downstream, when swirl no longer drives the wake's behaviour, it is expected that $\delta_{1/2}$ will also scale as $x^{1/2}$ (if non-equilibrium is still relevant), which was observed in the recent findings of \cite{lingkan2023assessment}.

\begin{figure}
                  \centering
                  \begin{subfigure}{0.8\textwidth}
                  \centerline{\includegraphics[width=\textwidth]{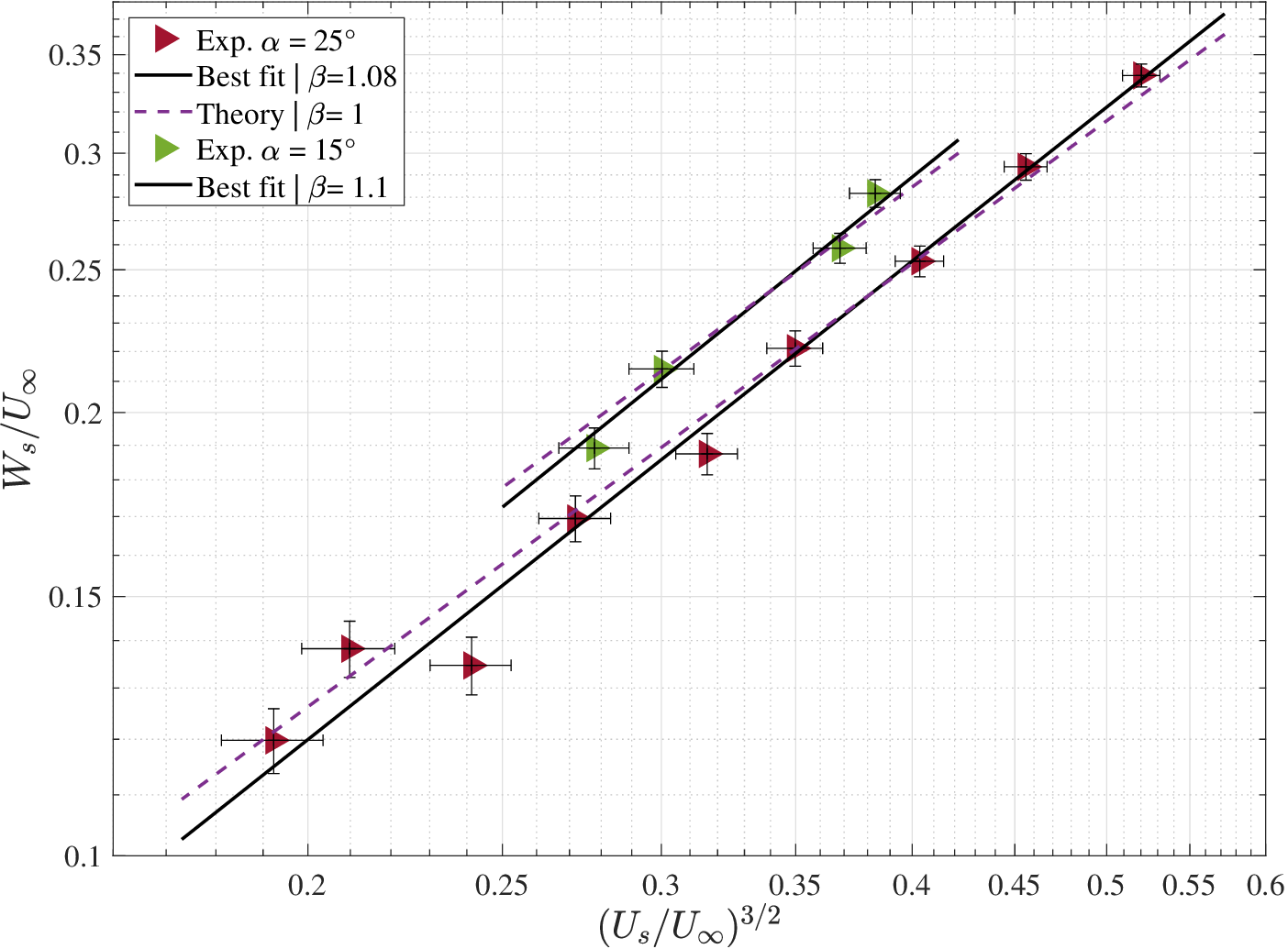}}
                  \end{subfigure}
                
                  \caption{Scaling plots of $W_s(x)$ vs. $U_s(x)^{3/2}$ for cases $\alpha=15\degree $ and $\alpha=25\degree $.}
                \label{fig:Ws_f_Usscaling}
                \end{figure}

The swirl decay  for cases $\alpha=15\degree $ and $\alpha=25\degree $ is emphasised in Fig. \ref{fig:Ws_scaling} and shows the downwind evolution of $W_s(x)/U_\infty$ along with their respective fitted scaling law. Once again, the data matches the non-equilibrium predictions ($\beta_{W} = -3/2$). Altogether, these results point out that the region of interest in the wake is well approximated by scaling laws based on non-equilibrium similarity. In particular, the proposed scaling law (\ref{eq:Ws_novel_scaling}) was verified in our data. The fact that the same conclusions are obtained for cases $\alpha=15\degree$ $\alpha=25\degree$ suggests that porosity may play a lesser role than swirl in the near field of the wake. Echoing the discussions made in the introduction, swirl might be the dominant initial condition of the wake. However, in order to fully confirm this claim, future studies featuring lower porosity actuator discs are needed. Moreover, as discussed in \S\ref{sec:similarity}, both similarity analyses result in the same scaling law between the swirl decay and the wake recovery, such as $W_s/U_\infty \sim (U_s/U_\infty)^{3/2}$. The linear fit plots of these quantities is reported in Fig. \ref{fig:Ws_f_Usscaling} for cases $\alpha=15\degree $ and $\alpha=25\degree $. It is found that the data supports this scaling law, especially for the $\alpha=25\degree$ case.

\subsection{Mean velocity deficit}

            \begin{figure}
              \centering
              \begin{subfigure}{0.9\textwidth}
              \includegraphics[width=\textwidth]{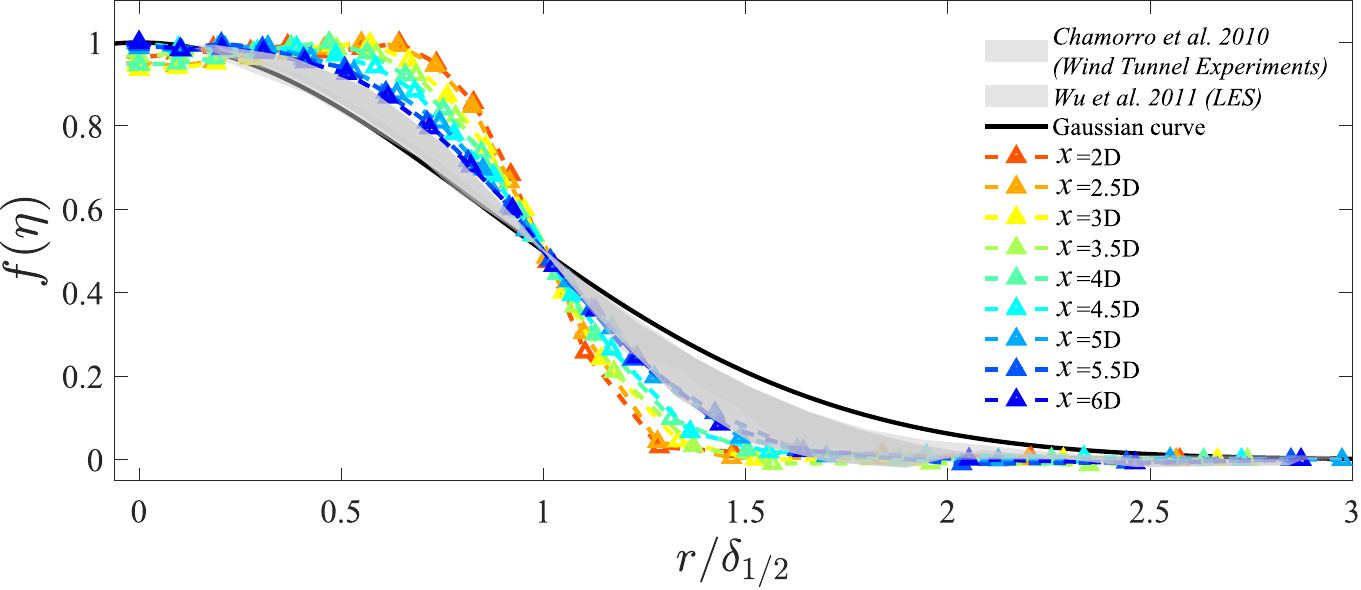}
              \subcaption{}
              \label{fig:self_similar_Ua}
              \end{subfigure}

              \centering
              \begin{subfigure}{0.9\textwidth}
              \includegraphics[width=\textwidth]{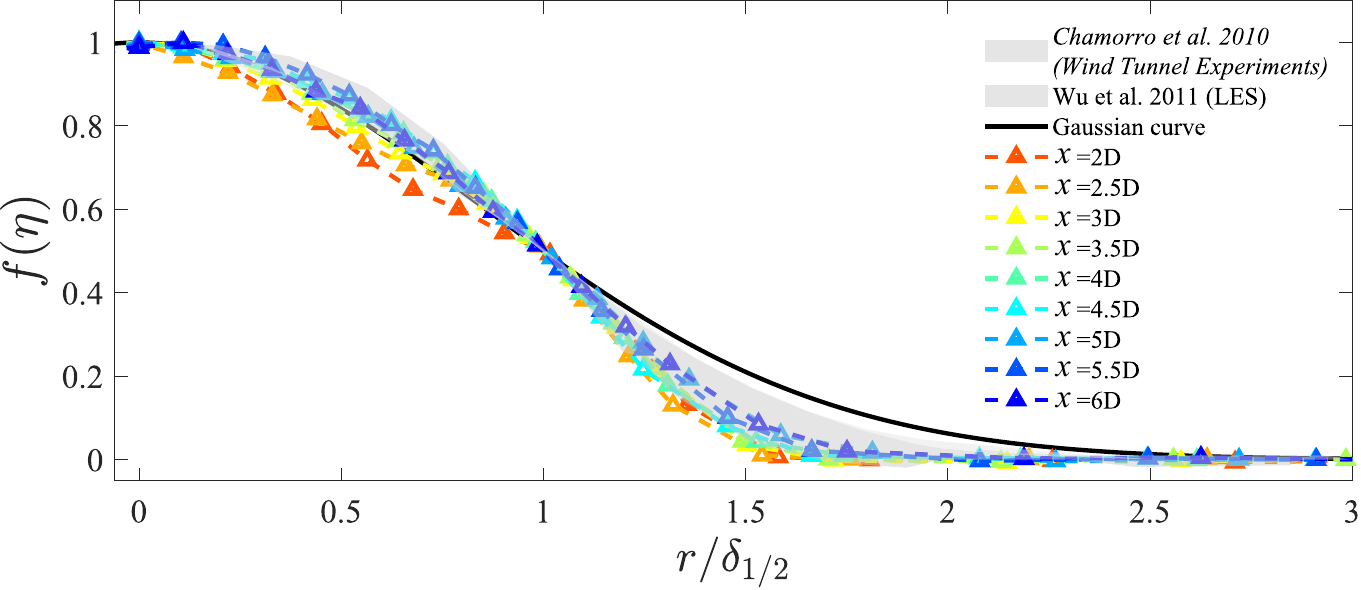}
              \subcaption{}
              \label{fig:self_similar_Ub}
              \end{subfigure}

              \centering
              \begin{subfigure}{0.9\textwidth}
              \includegraphics[width=\textwidth]{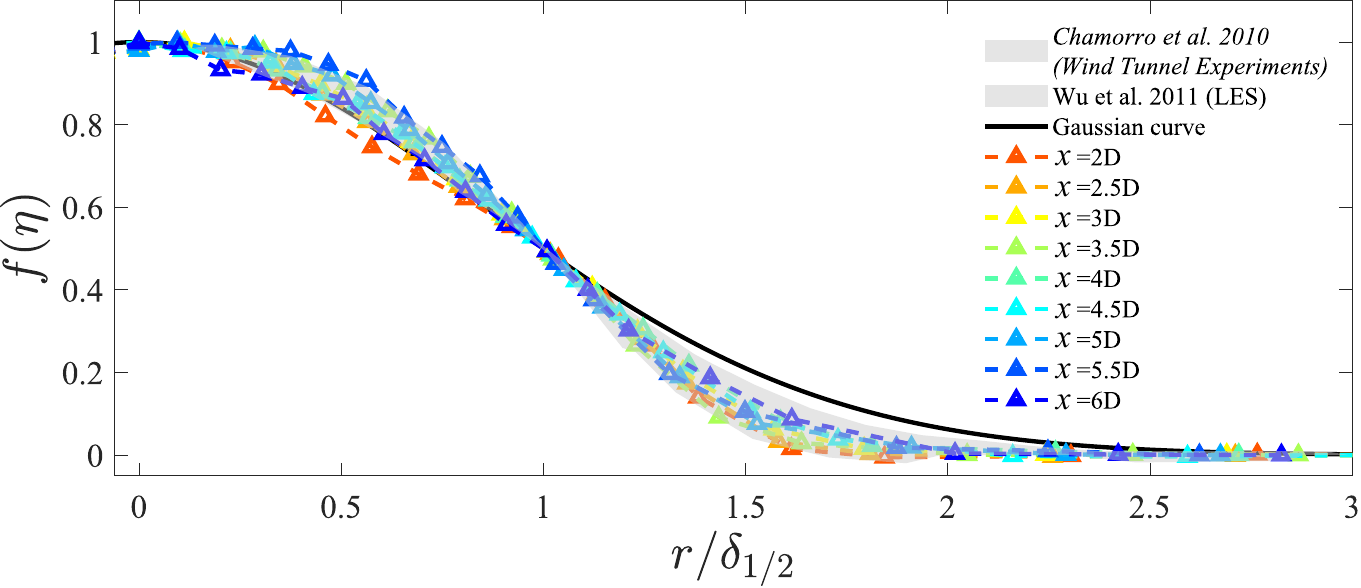}
              \subcaption{}
              \label{fig:self_similar_Uc}
              \end{subfigure}

              \caption{Self-similar streamwise velocity deficit profiles for (a) the $\alpha=0\degree $ case, (b) the $\alpha=15\degree $ case and for the (c) $\alpha=25\degree $ case. Black solid line: theoretical self-similar Gaussian curve, grey shaded area: experimental and numerical data reported in the literature.}
            \label{fig:self_similar_U}
            \end{figure}

\noindent A striking observation can be made regarding the scaling analysis of the generated wakes: the wake of the $\alpha=0\degree$ case did not scale with any known law. In particular, the mean velocity deficit appeared to recover very slowly and exhibited an axisymmetry defect. Both of these characteristics were corrected by the presence of swirl. Since the mean velocity deficit $\Delta U$ is central in wind turbine performance modelling \citep{PorteAgel2019}, the following section will examine how swirl modified the properties of this critical parameter. The profiles of $\Delta U$ are reported in Fig. \ref{fig:self_similar_U} using similarity variables. For comparison purposes, data reported for experimental lab-scale \citep{Chamorro2010} and numerical (LES) real-scale \citep{Wu2012} wind turbines are represented by the grey shaded area in Fig. \ref{fig:self_similar_U}. The Gaussian profile model proposed by \citet{Bastankhah2014} was in fact compared to these exact data sets and used as a validation criteria.

The reference case (figure \ref{fig:self_similar_Ua}), exhibits velocity profiles with a trilby hat shape up until $x^\star=6$ where the profiles only start to collapse towards a Gaussian shape. In contrast, the swirling wake streamwise velocity profiles (figures \ref{fig:self_similar_Ub} and \ref{fig:self_similar_Uc}) collapse and show self-similarity already at $x^\star=3$. Furthermore, the Gaussian profiles show good agreement with wind tunnel and numerical simulations reported in the literature for rotating wind turbines (grey shaded area). This evidences that one of the main mechanisms governing self-similarity in the wakes of wind turbines is the swirling motion of the wake. Similarly to the case of a swirling jet shown in the work of \cite{Shiri2008}, the mixing is enhanced and self-similarity is sped up by the swirling motion. Therefore, the proposed porous disc model better reproduces the physics in the near wake of a rotating wind turbine, in comparison to previous porous discs generating non-swirling wakes.

            \begin{figure}
              \centering
            
              \includegraphics[width=0.8\textwidth]{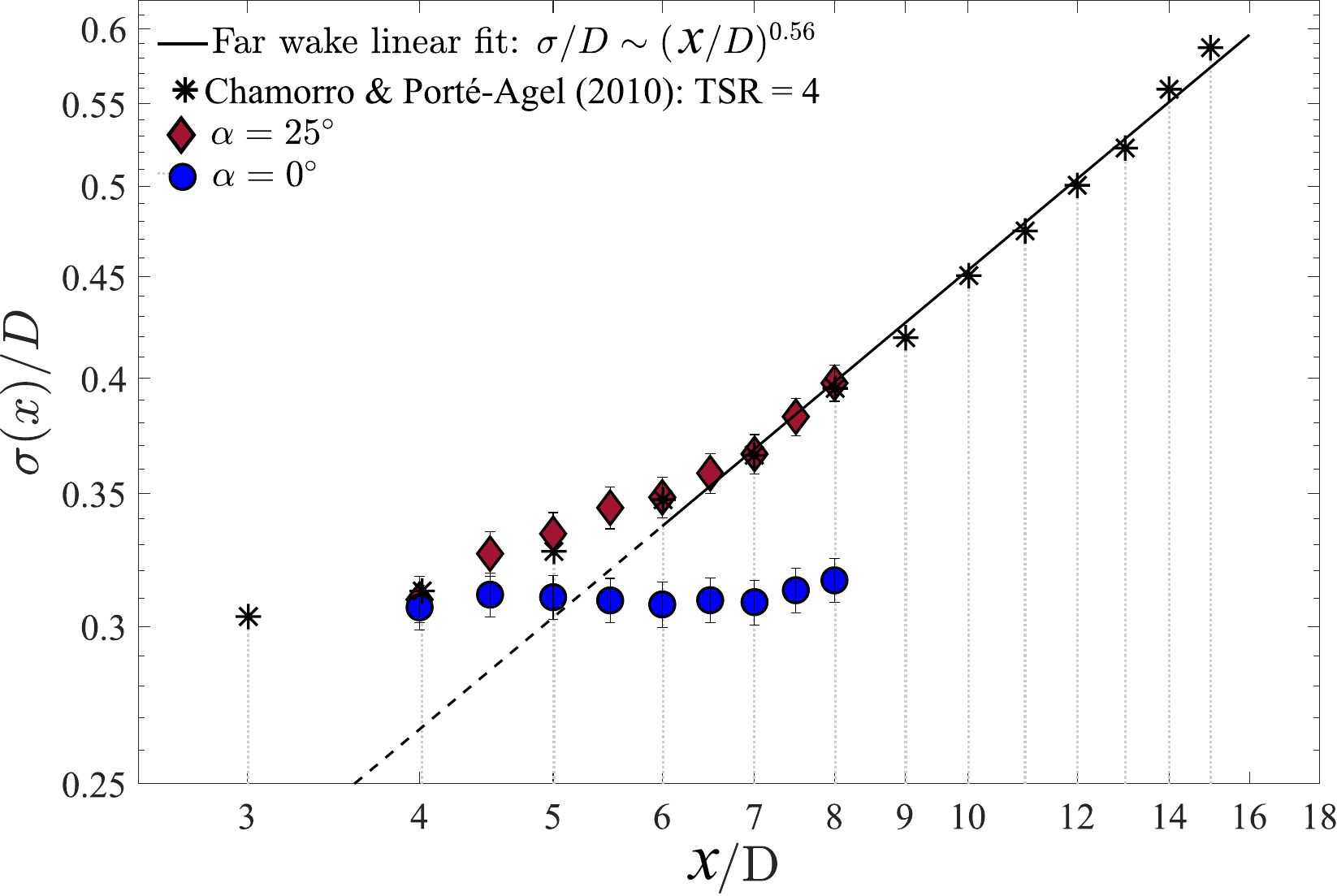}
              \caption{Normalised standard deviations of the streamwise velocity deficit profiles for cases $\alpha=0\degree $ and $\alpha=25\degree $ compared to the data reported in \cite{Chamorro2010}. Fitted linear trends are represented in solid lines.}
            \label{fig:PA_sigma}
            \end{figure}

            In order to compare the wake's width to the reported trends for rotating turbines, the wake width can be calculated as an equivalent standard deviation $\sigma(x)$. In the work of \cite{Bastankhah2014} for example, the proposed gaussian velocity deficit model was fitted to numerous data points also reported in this work. In particular, the data points reported in \cite{Chamorro2010} are shown in figure \ref{fig:PA_sigma}, where the wake of a lab-scale three-bladed rotating wind turbine immersed in a boundary layer was investigated. For our data, $\sigma(x)$ can be directly calculated as the square root of the calculated variance from the streamwise velocity deficit profiles. Figure \ref{fig:PA_sigma} compares the downwind evolution of $\sigma(x)$ along with the data reported in \cite{Chamorro2010}. The data points from the literature show a rupture in the trend of the wake expansion around $x^\star=6$. From $x^\star=2D$ to $x^\star=6D$ a light slope ($\approx x^{0.15}$) is observed which does not correspond to any known scaling laws. Beyond this point however, the trend abruptly changes and endorses a wake expansion law closer to the non-equilibrium scaling $\sigma \sim x^{1/2}$ \citep{dairay_non-equilibrium_2015} than to the classical equilibrium scaling $x^{1/3}$ \citep{Johansson2003}. Interestingly, the data obtained for $\alpha=25\degree $ mimics this behaviour. In the near wake, the swirling velocity will be the driving factor admitting a length scale related to swirl. The velocity deficit will take over as swirl decays, imposing its length scale $\sigma$. This is inline with what was theorised in section \S\ref{sec:similarity} regarding the existence of two scales governing the streamwise evolution of the swirling wake when $\hat{S} = \mathcal{O}(1)$ \citep{reynolds1962similarity}. On the other hand, the evolution of $\sigma(x)$ for case $\alpha=0\degree$ appears to remain roughly constant at all streamwise positions as this case has not reached a self-preserving state yet. These results underscore the central role that swirl plays in near wake development and the critical importance of incorporating it in future wind turbine studies based on the actuator disc model.

\section{Conclusions }

\noindent In this study, the absence of swirl in the wake of a porous disc wind turbine surrogate was tackled. The theoretical examination of the turbulent swirling wake produced by an actuator disc unveiled intriguing phenomena with the added ingredient of angular momentum conservation. Additional length and velocity scales of the flow were introduced in the problem thus enriching the framework of similarity theory as discussed by \cite{reynolds1962similarity}. In particular, it was shown that the conservation of angular momentum adds an extra degree of freedom to the similarity analysis where two scales coexist and can potentially drive the development of the wake depending on the swirl number dimensionless parameter $\hat{S}$. It was revealed that non-equilibrium turbulence is a more appropriate similarity framework for swirling flows since it avoids making assumptions about the swirl number asymptotic values. This scenario lead to the identification of alternative scaling laws supported by recent literature. Additionally, a novel non-equilibrium scaling law for swirl decay was proposed, expressed as $W_s/U_\infty \sim (x/\theta)^{-3/2}$. To validate our theoretical analysis, an extensive experimental investigation was conducted on the swirling turbulent wake generated by a modified porous disc now including swirl. 

A porous disc featuring non-uniform porosity $\beta$ was designed taking inspiration from the work of \cite{Camp2016} and \cite{helvig_2021}. To passively generate swirl, the blades of the porous disc were pitched of an angle $\alpha$. Two regimes were underscored based on the pitch angle $\alpha$: the attached flow regime and the stall regime. In the attached flow regime, swirl magnitude increases linearly and drag is constant. In the stall regime, swirl starts to decrease with $\alpha$ and drag increases linearly due to the decreased apparent porosity. A critical pitch angle was evidenced $\alpha_{c} \in [16\degree, 18\degree]$ beyond which stall occurs. Cases $\alpha=0\degree$, $\alpha=15\degree$ and $\alpha=25\degree$ were chosen as key pitch angles since the $\alpha=15\degree$ case generated a maximum amplitude of swirl at iso-porosity while the $\alpha=25\degree$ case generated comparable levels of swirl with a decreased apparent porosity. Swirl was fully characterised for these cases and showed swirl number values ($\hat{S} \approx 0.3$) comparable to what is reported for real wind turbines \citep{dufresne2013experimental,bortolotti2019iea,holmes2022impact}. It was shown that swirl generated a low pressure core at the centre of the wake near $(Y,Z) = (0,0)$. For the non-swirling wake ($\alpha=0\degree$), the low pressure area was located near the mast region, where a strong downwash effect was observed \citep{Aubrun2019}. It was found that swirl enhanced the wake's axisymmetry, resisting downwash. 
A scaling analysis revealed that the streamwise evolution of the wake's properties showed very good agreement with the non-equilibrium similarity predictions. In particular, the novel non-equilibrium swirl decay scaling law was obtained for $\alpha=15\degree$ and $\alpha=25\degree$. The universal scaling law coupling the characteristic axial velocity deficit and the characteristic swirling velocity $W_s(x)/U_\infty \sim (U_s(x)/U_\infty)^{3/2}$ was also found in a restricted range of streamwise locations. Identical results were obtained for cases $\alpha=15\degree$ and $\alpha=25\degree$ which suggested that swirl is the dominant initial condition of the wake relative to porosity. It is thus argued that swirl triggers self-preservation in the wake of the porous disc regardless of porosity. However, to firm up this conclusion, additional experiments need to be carried out with a lower porosity set of discs. Interestingly, none of the known scaling laws were found for the $\alpha=0\degree$ case in the region of interest. A close inspection of the mean velocity deficit profiles revealed that self-similar collapse was not yet reached for this case, although non-equilibrium turbulence applies. However, self-similar data collapse of the mean velocity deficit $U_s(x)/U_\infty$ profiles was enhanced by swirl,indicating that swirl governs the scaling laws for the mean wake properties. Self-similar gaussian profiles were obtained starting at $X=3D$ for the swirling wakes. These profiles were shown to better match the velocity deficit profiles from recent literature \citep{wu2013simulation,Chamorro2010} used as benchmark data to validate the widely known wake model proposed by \cite{Bastankhah2014}. Overall, this study proved the critical role of the swirling motion in shaping the near wake and governing its development. Our results underscore the importance of incorporating swirl in future wind turbine wake studies featuring the actuator disc model. Additionally, this work introduced a novel scaling law for the mean swirl decay based on non-equilibrium self-similarity which sets the stage for future wake models including swirl. Furthermore, we proposed an innovative porous disc design that passively introduces swirl in a cost-effective manner featuring no blockage alteration and swirl numbers comparable to those of real wind turbines under various operating conditions. 

\medskip

\noindent \textbf{Supplementary material.} Supplementary material is available at.

\medskip

\noindent \textbf{Acknowledgements.} We thank S. Loyer for his technical support during the experiments.

\medskip

\noindent \textbf{Declaration of interests.} The authors report no conflict of interest.

\bibliographystyle{jfm}

\bibliography{jfm}

\begin{thebibliography}{60}
\expandafter\ifx\csname natexlab\endcsname\relax\def\natexlab#1{#1}\fi
\def\au#1{#1} \def\ed#1{#1} \def\yr#1{#1}\def\at#1{#1}\def\jt#1{\textit{#1}}
  \def\bt#1{#1}\def\bvol#1{\textbf{#1}} \def\vol#1{#1} \def\pg#1{#1}
  \def\publ#1{#1}\def\arxiv#1{#1}\def\org#1{#1}\def\st#1{\textit{#1}}

\bibitem[Alekseenko {\em et~al.\/}(1999)Alekseenko, Kuibin, Leonidovich \&
  Shtork]{alekseenko1999helical}
{\sc \au{Alekseenko, S.~V.}, \au{Kuibin, P.~A.}, \au{Leonidovich, V. Okulov~S.}
  \& \au{Shtork, S.~I.}} \yr{1999}  \at{Helical vortices in swirl flow}.
  \jt{Journal of Fluid Mechanics}  \bvol{382},  \pg{195--243}.

\bibitem[Anderson(2011)]{anderson2011fundamentals}
{\sc \au{Anderson, J.~D.}} \yr{2011} {\em Fundamentals of aerodynamics\/}.
  \publ{McGraw-Hill}.

\bibitem[Aubrun(2013)]{Aubrun2013}
{\sc \au{Aubrun, S.}} \yr{2013}  \at{Wind turbine wake properties: Comparison
  between a non-rotating simplified wind turbine model and a rotating model}.
  \jt{J. Wind Eng. Ind. Aerodyn.} .

\bibitem[Aubrun {\em et~al.\/}(2019)Aubrun, Bastankhah, Cal, Conan, Hearst,
  Hoek \& et~al.]{Aubrun2019}
{\sc \au{Aubrun, S.}, \au{Bastankhah, M.}, \au{Cal, R.~B.}, \au{Conan, B.},
  \au{Hearst, R.J}, \au{Hoek, D.} \& \au{et~al., A.~Zasso}} \yr{2019}
  \at{Round-robin tests of porous disc models}.  \jt{Journal of Physics:
  Conference Series}  \bvol{47}.

\bibitem[Bastankhah \& Porté-Agel(2014)]{Bastankhah2014}
{\sc \au{Bastankhah, M.} \& \au{Porté-Agel, F.}} \yr{2014}  \at{A new
  analytical model for wind-turbine wakes}.  \jt{Renewable Energy}  \bvol{70}.

\bibitem[Bevilaqua \& Lykoudis(1978)]{bevilaqua1978turbulence}
{\sc \au{Bevilaqua, P.~M.} \& \au{Lykoudis, P.~S.}} \yr{1978}  \at{Turbulence
  memory in self-preserving wakes}.  \jt{Journal of Fluid Mechanics}
  \bvol{89}~(3),  \pg{589--606}.

\bibitem[Bortolotti {\em et~al.\/}(2019)Bortolotti, Tarres, Dykes, Merz,
  Sethuraman, Verelst \& Zahle]{bortolotti2019iea}
{\sc \au{Bortolotti, P.}, \au{Tarres, H.~C.}, \au{Dykes, K.}, \au{Merz, K.},
  \au{Sethuraman, L.}, \au{Verelst, D.} \& \au{Zahle, F.}} \yr{2019}  \at{Iea
  wind tcp task 37: Systems engineering in wind energy-wp2. 1 reference wind
  turbines} .

\bibitem[Bossuyt {\em et~al.\/}(2017)Bossuyt, Meneveau \&
  Meyers]{bossuyt2017wind}
{\sc \au{Bossuyt, Juliaan}, \au{Meneveau, Charles} \& \au{Meyers, Johan}}
  \yr{2017}  \at{Wind farm power fluctuations and spatial sampling of turbulent
  boundary layers}.  \jt{Journal of Fluid Mechanics}  \bvol{823},
  \pg{329--344}.

\bibitem[Bruun(1996)]{bruun1996hot}
{\sc \au{Bruun, H.~H.}} \yr{1996} {\em Hot-wire anemometry: principles and
  signal analysis\/}.  \publ{Oxford University Press}.

\bibitem[Camp \& Cal(2016)]{Camp2016}
{\sc \au{Camp, E.~H.} \& \au{Cal, R.~B.}} \yr{2016}  \at{Mean kinetic energy
  transport and event classification in a model wind turbine array versus an
  array of porous disks: Energy budget and octant analysis}.  \jt{Physical
  Review Fluids}  \bvol{1}.

\bibitem[Castro(1971)]{castro1971perforatedplates}
{\sc \au{Castro, I.~P}} \yr{1971}  \at{Wake characteristics of two-dimensional
  perforated plates normal to an air-stream}.  \jt{Journal of Fluid Mechanics}
  \bvol{46},  \pg{599--609}.

\bibitem[Chamorro \& Porte-Agel(2011)]{chamorro2011turbulent}
{\sc \au{Chamorro, L.~P.} \& \au{Porte-Agel, F.}} \yr{2011}  \at{Turbulent flow
  inside and above a wind farm: a wind-tunnel study}.  \jt{Energies}
  \bvol{4}~(11),  \pg{1916--1936}.

\bibitem[Chamorro \& Porté-Agel(2010)]{Chamorro2010}
{\sc \au{Chamorro, L.~P.} \& \au{Porté-Agel, F.}} \yr{2010}  \at{Effects of
  thermal stability and incoming boundary-layer flow characteristics on
  wind-turbine wakes: A wind-tunnel study}.  \jt{Boundary-Layer Meteorology}
  \bvol{136}.

\bibitem[Chernykh {\em et~al.\/}(2005)Chernykh, Demenkov \&
  Kostomakha]{chernykh2005swirling}
{\sc \au{Chernykh, G.~G.}, \au{Demenkov, A.~G.} \& \au{Kostomakha, V.~A.}}
  \yr{2005}  \at{Swirling turbulent wake behind a self-propelled body}.
  \jt{International Journal of Computational Fluid Dynamics}  \bvol{19}~(5),
  \pg{399--408}.

\bibitem[Dairay {\em et~al.\/}(2015)Dairay, Obligado \&
  Vassilicos]{dairay_non-equilibrium_2015}
{\sc \au{Dairay, T.}, \au{Obligado, M.} \& \au{Vassilicos, J.C}} \yr{2015}
  \at{Non-equilibrium scaling laws in axisymmetric turbulent wakes}.  \jt{J.
  Fluid Mech.}  \bvol{781},  \pg{166--195}.

\bibitem[Duffman(1980)]{huffman1980calibration}
{\sc \au{Duffman, G.~D.}} \yr{1980}  \at{Calibration of triaxial hot-wire
  probes using a numerical search algorithm}.  \jt{Journal of Physics E:
  Scientific Instruments}  \bvol{13}~(11),  \pg{1177}.

\bibitem[Dufresne(2013)]{dufresne2013experimental}
{\sc \au{Dufresne, N.~P.}} \yr{2013}  \at{Experimental investigation of the
  turbulent axisymmetric wake with rotation generated by a wind turbine}. PhD
  thesis, University of New Hampshire.

\bibitem[Frandsen {\em et~al.\/}(2006)Frandsen, , Barthelmie, Pryor, Rathmann,
  Larsen, H{\o}jstrup \& Th{\o}gersen]{frandsen2006analytical}
{\sc \au{Frandsen, S.}, , \au{Barthelmie, R.~J.}, \au{Pryor, S.}, \au{Rathmann,
  O.}, \au{Larsen, S.}, \au{H{\o}jstrup, J.} \& \au{Th{\o}gersen, M.}}
  \yr{2006}  \at{Analytical modelling of wind speed deficit in large offshore
  wind farms}.  \jt{Wind Energy: An International Journal for Progress and
  Applications in Wind Power Conversion Technology}  \bvol{9}~(1-2),
  \pg{39--53}.

\bibitem[George(1989)]{George89}
{\sc \au{George, W.~K.}} \yr{1989}  \at{The self-preservation of turbulent
  flows and its relation to initial conditions and coherent structures}.
  \jt{Advances in Turbulence} .

\bibitem[Helvig {\em et~al.\/}(2021)Helvig, Vinnes, Segalini, Worth \&
  Hearst]{helvig_2021}
{\sc \au{Helvig, S.~J.}, \au{Vinnes, K.~M.}, \au{Segalini, A.}, \au{Worth,
  A.~N.} \& \au{Hearst, R.J}} \yr{2021}  \at{A comparison of lab-scale free
  rotating wind turbines and actuator disks}.  \jt{Journal of Wind Engineering
  and Industrial Aerodynamics, American Society of Mechanical Engineers ASME}
  \bvol{209}.

\bibitem[Holmes \& Naughton(2022)]{holmes2022impact}
{\sc \au{Holmes, M.~J.} \& \au{Naughton, J.~M.}} \yr{2022}  \at{The impact of
  swirl and wake strength on turbulent axisymmetric wake evolution}.
  \jt{Physics of Fluids}  \bvol{34}~(9),  \pg{095101}.

\bibitem[Howland {\em et~al.\/}(2016)Howland, Bossuyt, Martinez-Tossas, Meyers
  \& Maneveau]{Howland2016}
{\sc \au{Howland, M.}, \au{Bossuyt, J.}, \au{Martinez-Tossas, L.~A.},
  \au{Meyers, J.} \& \au{Maneveau, C.}} \yr{2016}  \at{Wake structure in
  actuator disk models of wind turbines in yaw under uniform inflow
  conditions}.  \jt{Journal of Renewable and Sustainable Energy}  \bvol{8}.

\bibitem[Jensen(1983)]{jensen1983note}
{\sc \au{Jensen, N.O.}} \yr{1983}  \at{A note on wind turbine interaction}.
  \jt{Riso-M-2411, Risoe National Laboratory, Roskilde, Denmark}  \pg{p.~16}.

\bibitem[Johansson {\em et~al.\/}(2003)Johansson, George \&
  Gourlay]{Johansson2003}
{\sc \au{Johansson, P. B.~V.}, \au{George, W.~K.} \& \au{Gourlay, M.~J.}}
  \yr{2003}  \at{Equilibrium similarity, effects of initial conditions and
  local reynolds number on the axisymmetric wake}.  \jt{Phys. Fluids}
  \bvol{15}.

\bibitem[Joukowsky(1912)]{joukowsky1912vortex}
{\sc \au{Joukowsky, N.~E.}} \yr{1912}  \at{Vortex theory of screw propeller}.
  \jt{Trudy Otdeleniya Fizicheskikh Nauk Obshchestva Lubitelei Estestvoznaniya
  (in Russian)}  \bvol{16}~(1),  \pg{1--31}.

\bibitem[Kallio \& Stock(1992)]{kallio1992interaction}
{\sc \au{Kallio, G.~A.} \& \au{Stock, D.~E.}} \yr{1992}  \at{Interaction of
  electrostatic and fluid dynamic fields in wire—plate electrostatic
  precipitators}.  \jt{Journal of Fluid Mechanics}  \bvol{240},  \pg{133--166}.

\bibitem[Lee {\em et~al.\/}(2020)Lee, Kim, Khosronejad \&
  Kang]{lee2020experimental}
{\sc \au{Lee, J.}, \au{Kim, Y.}, \au{Khosronejad, A.} \& \au{Kang, S.}}
  \yr{2020}  \at{Experimental study of the wake characteristics of an axial
  flow hydrokinetic turbine at different tip speed ratios}.  \jt{Ocean
  Engineering}  \bvol{196},  \pg{106777}.

\bibitem[Li {\em et~al.\/}(2021)Li, Zhao, Liu \& Carmeliet]{Li2021}
{\sc \au{Li, H.}, \au{Zhao, Y.}, \au{Liu, J.} \& \au{Carmeliet, J.}} \yr{2021}
  \at{Physics-based stitching of multi-fov piv measurements for urban wind
  fields}.  \jt{Building and Environment}  \bvol{205}.

\bibitem[Lignarolo {\em et~al.\/}(2016)Lignarolo, Ragni \&
  Ferreira]{Lignarolo2016exp}
{\sc \au{Lignarolo, L. E.~M.}, \au{Ragni, D.} \& \au{Ferreira, C.~J.}}
  \yr{2016}  \at{Experimental comparison of a wind-turbine and of an
  actuator-disc near wake}.  \jt{J. Renewable Sustainable Energy} .

\bibitem[Lingkan \& Buxton(2023)]{lingkan2023assessment}
{\sc \au{Lingkan, E.~H.} \& \au{Buxton, O.}} \yr{2023}  \at{An assessment of
  the scalings for the streamwise evolution of turbulent quantities in wakes
  produced by porous objects}.  \jt{Renewable Energy} .

\bibitem[Masri {\em et~al.\/}(2004)Masri, Kalt \&
  Barlow]{masri2004compositional}
{\sc \au{Masri, A.~R.}, \au{Kalt, P.} \& \au{Barlow, R.~S.}} \yr{2004}  \at{The
  compositional structure of swirl-stabilised turbulent nonpremixed flames}.
  \jt{Combustion and Flame}  \bvol{137}~(1-2),  \pg{1--37}.

\bibitem[Mohebi {\em et~al.\/}(2017)Mohebi, Wood \&
  Martinuzzi]{mohebi2017turbulence}
{\sc \au{Mohebi, M.}, \au{Wood, D.} \& \au{Martinuzzi, R.~J.}} \yr{2017}
  \at{The turbulence structure of the wake of a thin flat plate at post-stall
  angles of attack}.  \jt{Experiments in Fluids}  \bvol{58},  \pg{1--18}.

\bibitem[Moisy {\em et~al.\/}(2011)Moisy, Morize, Rabaud \&
  Sommeria]{moisy2011decay}
{\sc \au{Moisy, F.}, \au{Morize, C.}, \au{Rabaud, M.} \& \au{Sommeria, J.}}
  \yr{2011}  \at{Decay laws, anisotropy and cyclone--anticyclone asymmetry in
  decaying rotating turbulence}.  \jt{Journal of Fluid Mechanics}  \bvol{666},
  \pg{5--35}.

\bibitem[Mora {\em et~al.\/}(2019)Mora, Pladellorens, Turr{\'o}, Lagauzere \&
  Obligado]{mora2019energy}
{\sc \au{Mora, D.~O.}, \au{Pladellorens, E.~M.}, \au{Turr{\'o}, P.~R.},
  \au{Lagauzere, M.} \& \au{Obligado, M.}} \yr{2019}  \at{Energy cascades in
  active-grid-generated turbulent flows}.  \jt{Physical Review Fluids}
  \bvol{4}~(10),  \pg{104601}.

\bibitem[Morris {\em et~al.\/}(2016)Morris, O’doherty, Mason-Jones \&
  O’Doherty]{morris2016evaluation}
{\sc \au{Morris, C.~E.}, \au{O’doherty, D.~M.}, \au{Mason-Jones, A.} \&
  \au{O’Doherty, T.}} \yr{2016}  \at{Evaluation of the swirl characteristics
  of a tidal stream turbine wake}.  \jt{International Journal of Marine Energy}
   \bvol{14},  \pg{198--214}.

\bibitem[Nakayama(1988)]{nakayama1988visualized}
{\sc \au{Nakayama, Y.}} \yr{1988} {\em Visualized flow: fluid motion in basic
  and engineering situations revealed by flow visualization\/}.  \publ{Pergamon
  Press}.

\bibitem[Nedic(2013)]{nedic2013fractal}
{\sc \au{Nedic, J.}} \yr{2013}  \at{Fractal-generated wakes}. PhD thesis,
  Imperial College of London.

\bibitem[Okulov {\em et~al.\/}(2015)Okulov, S{\o}rensen \&
  Wood]{okulov2015rotor}
{\sc \au{Okulov, V.~L.}, \au{S{\o}rensen, J.~N.} \& \au{Wood, D.~H.}} \yr{2015}
   \at{The rotor theories by professor joukowsky: Vortex theories}.
  \jt{Progress in aerospace sciences}  \bvol{73},  \pg{19--46}.

\bibitem[Pope(2000)]{Pope2000}
{\sc \au{Pope, S.~B.}} \yr{2000} {\em {Turbulent flows}\/}.  \publ{Cambridge
  University Press}.

\bibitem[Porté-Agel(2019)]{PorteAgel2019}
{\sc \au{Porté-Agel, F.}} \yr{2019}  \at{Wind-turbine and wind-farm flows: A
  review}.  \jt{Springer} .

\bibitem[Raffel {\em et~al.\/}(2007)Raffel, Willert \& Kompenhans]{Raffel2007}
{\sc \au{Raffel, M.}, \au{Willert, C.~E.} \& \au{Kompenhans, J.}} \yr{2007}
  {\em Particle image velocimetry: a practical guide\/}.  \publ{Springer
  Science \& Business Media}.

\bibitem[Rankine(1865)]{rankine1865mechanical}
{\sc \au{Rankine, William John~Macquorn}} \yr{1865}  \at{On the mechanical
  principles of the action of propellers}.  \jt{Transactions of the Institution
  of Naval Architects}  \bvol{6}.

\bibitem[Reynolds(1962)]{reynolds1962similarity}
{\sc \au{Reynolds, A.~J.}} \yr{1962}  \at{Similarity in swirling wakes and
  jets}.  \jt{Journal of Fluid Mechanics}  \bvol{14}~(2),  \pg{241--243}.

\bibitem[Sforza {\em et~al.\/}(1981)Sforza, Sheerin \& Smorto]{sforza1981three}
{\sc \au{Sforza, P.~M.}, \au{Sheerin, P.} \& \au{Smorto, M.}} \yr{1981}
  \at{Three-dimensional wakes of simulated wind turbines}.  \jt{AIAA journal}
  \bvol{19}~(9),  \pg{1101--1107}.

\bibitem[Shanmughan {\em et~al.\/}(2020)Shanmughan, Passaggia, Mazellier \&
  Kourta]{shanmughan2020optimal}
{\sc \au{Shanmughan, R.}, \au{Passaggia, P.~Y}, \au{Mazellier, N.} \&
  \au{Kourta, A.}} \yr{2020}  \at{Optimal pressure reconstruction based on
  planar particle image velocimetry and sparse sensor measurements}.
  \jt{Experiments in Fluids}  \bvol{61},  \pg{1--19}.

\bibitem[Shiri(2010)]{Shiri2010}
{\sc \au{Shiri, A.}} \yr{2010}  \at{Turbulence measurements in a natural
  convection boundary layer and a swirling jet}. PhD thesis, Chalmers
  University of Technology, Göteborg, Sweden.

\bibitem[Shiri {\em et~al.\/}(2008)Shiri, George \& Naughton]{Shiri2008}
{\sc \au{Shiri, A.}, \au{George, W.~K.} \& \au{Naughton, J.~W.}} \yr{2008}
  \at{Experimental study of the far field of incompressible swirling jets}.
  \jt{AIAA Journal}  \bvol{46}.

\bibitem[Sreenivasan {\em et~al.\/}(1983)Sreenivasan, Prabhu \&
  Narasimha]{sreenivasan1983zero}
{\sc \au{Sreenivasan, K.~R.}, \au{Prabhu, A.} \& \au{Narasimha, R.}} \yr{1983}
  \at{Zero-crossings in turbulent signals}.  \jt{Journal of Fluid Mechanics}
  \bvol{137},  \pg{251--272}.

\bibitem[Steiros \& Hultmark(2018)]{steiros2018drag}
{\sc \au{Steiros, K.} \& \au{Hultmark, M.}} \yr{2018}  \at{Drag on flat plates
  of arbitrary porosity}.  \jt{Journal of Fluid Mechanics}  \bvol{853},
  \pg{R3}.

\bibitem[Stevens {\em et~al.\/}(2018)Stevens, Mart{\'\i}nez-Tossas \&
  Meneveau]{stevens2018comparison}
{\sc \au{Stevens, R. J. A.~M.}, \au{Mart{\'\i}nez-Tossas, L.~A.} \&
  \au{Meneveau, C.}} \yr{2018}  \at{Comparison of wind farm large eddy
  simulations using actuator disk and actuator line models with wind tunnel
  experiments}.  \jt{Renewable energy}  \bvol{116},  \pg{470--478}.

\bibitem[Townsend(1976)]{Townsend1976}
{\sc \au{Townsend, A.~A.}} \yr{1976} {\em The Structure of Turbulent shear
  flow, 2nd ed.\/}.  \publ{Cambridge University Press}.

\bibitem[Van~Kuik {\em et~al.\/}(2015)Van~Kuik, S{\o}rensen \&
  Okulov]{van2015rotor}
{\sc \au{Van~Kuik, GAM}, \au{S{\o}rensen, Jens~N{\o}rk{\ae}r} \& \au{Okulov,
  VL}} \yr{2015}  \at{Rotor theories by professor joukowsky: momentum
  theories}.  \jt{Progress in Aerospace Sciences}  \bvol{73},  \pg{1--18}.

\bibitem[Vassilicos(2015)]{vassilicos2015dissipation}
{\sc \au{Vassilicos, J.~C.}} \yr{2015}  \at{Dissipation in turbulent flows}.
  \jt{Annual review of fluid mechanics}  \bvol{47},  \pg{95--114}.

\bibitem[Vincent(2007)]{vincent2007aerosol}
{\sc \au{Vincent, J.~H.}} \yr{2007} {\em Aerosol sampling: science, standards,
  instrumentation and applications\/}.  \publ{John Wiley \& Sons}.

\bibitem[Vinnes(2023)]{vinnes2023actuator}
{\sc \au{Vinnes, M.~K.}} \yr{2023}  \at{The actuator disk as a wind turbine
  model: An experimental assessment of the fluid dynamics}. PhD thesis,
  Norwegian University of Science and Technology.

\bibitem[Wieneke(2015)]{Wieneke2015PIV}
{\sc \au{Wieneke, B.}} \yr{2015}  \at{Piv uncertainty quantification from
  correlation statistics}.  \jt{Measurement Science and Technology}  \bvol{26}.

\bibitem[Wosnik \& Dufresne(2013)]{Wosnik2013}
{\sc \au{Wosnik, M.} \& \au{Dufresne, N.}} \yr{2013}  \at{Experimental
  investigation and similarity solution of the axisymmetric turbulent wake with
  rotation}.  \jt{Fundamental Issues and Perspectives in Fluid Mechanics, ASME
  2013 Fluids Engineering Division Summer Meeting}  \bvol{1B}.

\bibitem[Wu \& Porté-Agel(2012)]{Wu2012}
{\sc \au{Wu, Y.~T.} \& \au{Porté-Agel, F.}} \yr{2012}  \at{Atmospheric
  turbulence effects on wind-turbine wakes: An les study}.  \jt{Energies}
  \bvol{5}.

\bibitem[Wu \& Porté-Agel(2013)]{wu2013simulation}
{\sc \au{Wu, Y.~T.} \& \au{Porté-Agel, F.}} \yr{2013}  \at{Simulation of
  turbulent flow inside and above wind farms: model validation and layout
  effects}.  \jt{Boundary-layer meteorology}  \bvol{146},  \pg{181--205}.

\bibitem[Wygnanski {\em et~al.\/}(1986)Wygnanski, Champagne \&
  Marasli]{wygnanski1986large}
{\sc \au{Wygnanski, I}, \au{Champagne, F} \& \au{Marasli, B}} \yr{1986}  \at{On
  the large-scale structures in two-dimensional, small-deficit, turbulent
  wakes}.  \jt{Journal of Fluid Mechanics}  \bvol{168},  \pg{31--71}.

\end{thebibliography}

\end{document}